\documentclass[journal,twocolumn,10pt]{IEEEtranTCOM}
\usepackage[pdftex]{graphicx,xcolor}
\usepackage[fleqn]{amsmath}
\usepackage{newtxtext}
\usepackage[varg]{newtxmath}
\usepackage{graphicx, epstopdf,cite,subfigure,cite}
\usepackage{amsfonts}
\usepackage[cmex10]{mathtools}
\usepackage{multirow}
\usepackage{url}
\usepackage{graphics}
\usepackage{array}
\usepackage{color,soul}
\usepackage{url, cite}
\usepackage{bbold}
\usepackage{float}
\usepackage{array}
\usepackage{adjustbox}

\newtheorem{theorem}{\bf Theorem}

\newtheorem{lemma}{{\bf Lemma}}

\newcolumntype{L}{>{\centering\arraybackslash}m{3cm}}
\normalsize

\begin{document}

\title{TRIDENT: A load-balancing Clos-network Packet Switch with Queues between Input and Central Stages and In-Order Forwarding}

\author{Oladele~Theophilus Sule and Roberto~Rojas-Cessa,~\IEEEmembership{Senior Member,~IEEE} 
	\thanks{O.T. Sule and R. Rojas-Cessa are with the Department of Electrical and Computer Engineering, New Jersey Institute of Technology, Newark, NJ 07102. Email: \{{ots5, rojas}\}@njit.edu.

\noindent \textit{(Corresponding author: Oladele Theophilus Sule)}}}


\maketitle

\begin{abstract}
We propose a three-stage load balancing packet switch and its configuration scheme. The input- and central-stage switches are bufferless crossbars, and the output-stage switches are buffered crossbars. We call this switch ThRee-stage Clos-network swItch with queues at the middle stage and DEtermiNisTic scheduling (TRIDENT), and the switch is cell based. The proposed configuration scheme uses predetermined and periodic interconnection patterns in the input and central modules to load-balance and route traffic, therefore; it has low configuration complexity. The operation of the switch includes a mechanism applied at input and output modules to forward cells in sequence. 
TRIDENT achieves 100\% throughput under uniform and nonuniform admissible traffic with independent and identical distributions (i.i.d.). The switch achieves this high performance using a low-complexity architecture while performing in-sequence forwarding and no central-stage expansion or memory speedup. Our discussion includes throughput analysis, where we describe the operations the configuration mechanism performs on the traffic traversing the switch, and proof of in-sequence forwarding. We present a simulation analysis as a practical demonstration of the switch performance under uniform and nonuniform i.i.d. traffic.
\end{abstract}


\IEEEpeerreviewmaketitle

\section{Introduction}
Clos networks are very attractive for building large-size switches \cite{clos1953study}. Most Clos-network switches adopt three stages, where each stage uses switch modules as building blocks. The modules of the first, second, and third stages are called input, central, and output modules, and they are denoted as IM, CM, and OM, respectively. Overall, Clos-network switches require fewer switching units (crosspoint elements), than a single-stage switch of equivalent size, and thus may require less building hardware. 
The hardware reduction of a Clos-network switch often increases its configuration complexity. 
In general, a Clos-network switch requires configuring its modules before forwarding packets through them.

We consider for the remainder of this paper that the proposed packet switch is cell-based; this is, upon arrival in an input port of a switch, packets of variable size are segmented into fixed-size cells and re-assembled at the output port, after being switched through the switch. The smallest size of a cell depends on the response time of the fabric and reconfiguration time.

Clos-network switches can be categorized based on whether a stage performs space- (S) or memory-based (M) switching into SSS (or S$^3$) \cite{lee1997path,chao2003matching}, MSM \cite{atlanta, oki2009analysis,kleban2013static,kleban2007modified,rojas2004maximum},  MMM \cite{chao2005trueway, chrysos2006scheduling, dong2011non, xia2016practical,hassen2017high}, SMM \cite{li2005space}, and SSM \cite{lin2013minimizing,rojas2006scalable},  among the most popular ones. Compared to the other categories, S$^3$ switches require the smallest amount of hardware, but their configuration complexity is high. Despite having a reduced configuration time, MMM switches, must deal with internal blocking and the multiplicity of input-output paths associated with diverse queuing delays \cite{chao2005trueway,dong2012mcs}. In general, switches with buffers in either the central or output stage are prone to forwarding packets out of sequence because of variable queue lengths, making in-sequence transmission mechanisms or re-sequencing a required feature. 

Traffic load balancing is a technique that improves the performance of switching and reduces the configuration complexity \cite{chang2002load}. Such a technique is especially attractive for its application to Clos-network switches as these suffer from high configuration complexity. A large number of network applications such as those used in network virtualization and data center network, adopt load balancing techniques to obtain high performance \cite{multipath-switching-slice12,lb-virtual-topologies,dixit2013impact}. Load balancing finds its application in wireless networks \cite{wang2009energy, le2011multipath, ploumidis2017flow}.

Predetermined and periodic permutations scheduling mechanism may be used for load-balancing and routing to achieve high switching performance\cite{chao2005trueway, zhang2014space, sule2018split}. A switch using a deterministic and periodic schedule may require queues between the load-balancing and routing stages. These queues store the cells while they wait for forwarding. These queues enable multiple interconnection paths between the load-balancing stage and the other stages of the switch, but they also make these switches prone to forwarding cells out of sequence \cite{chang2002load}. Re-sequencing \cite{changload} and out-of-sequence prevention mechanisms \cite{keslassy2002maintaining,rojas2016interconnections}, as they become switch components, may affect the switching performance and increase complexity.

The issues above raise the question, can a load-balancing Clos-network switch attain high switching performance, low configuration complexity, and in-sequence cell forwarding without resorting to memory speedup nor switch expansion?

We answer this question affirmatively in this paper by proposing a load-balancing Clos-network switch that has buffers placed between the IMs and CMs. Furthermore, we use OMs implemented with buffered crossbars with per-flow queues. The switch is called ThRee-stage Clos swItch with queues at the middle stage and DEtermiNisTic scheduling (TRIDENT). This switch uses predetermined and periodic interconnection patterns for the configuration of IMs and CMs. The incoming traffic is load-balanced by IMs and routed by CMs and OMs. The result is a switch that attains high throughput under admissible traffic with independent and identical distribution (i.i.d.) and uses a configuration scheme with $O(1)$ complexity. The switch also adopts an in-sequence forwarding mechanism at the input ports and output modules to keep cells in sequence.

The motivation for adopting this configuration method is its simplicity and low complexity.
For instance, TRIDENT reduces the amount of hardware needed by another load balancing switch \cite{sule2018split} and it also reduces the complexity of the in-sequence mechanism. The configuration approach used by TRIDENT also provides full utilization of the switch fabric and requires a small configuration time because of its predeterministic and periodic pattern. Our solution overcomes the required module or port matching, which are complex and time consuming, as required by other schemes.

We analyze the performance of the proposed switch by modeling the effect of each stage on the traffic passing through the switch. In addition, we study the performance of the switch through traffic analysis and by computer simulation. We show that the switch attains 100\% throughput under several admissible traffic models, including traffic with uniform and nonuniform distributions, and demonstrate that the switch forwards cells to the output ports in sequence. This high switching performance is achieved without resorting to speedup nor switch expansion.

The remainder of this paper is organized as follows: Section \ref{sec:switch-architecture} introduces the TRIDENT switch. Section \ref{sec:throughput-analysis} presents the throughput analysis of the proposed switch.  Section \ref{sec:proof} presents a proof of the in-sequence forwarding property of TRIDENT. Section \ref{sec:simulation-performance} presents a simulation study on the performance of the proposed switch. Section \ref{sec:conclusions} presents our conclusions.

\section{Switch Architecture}
\label{sec:switch-architecture}
\begin{figure*}[htb]
	\includegraphics[width=6.5in]{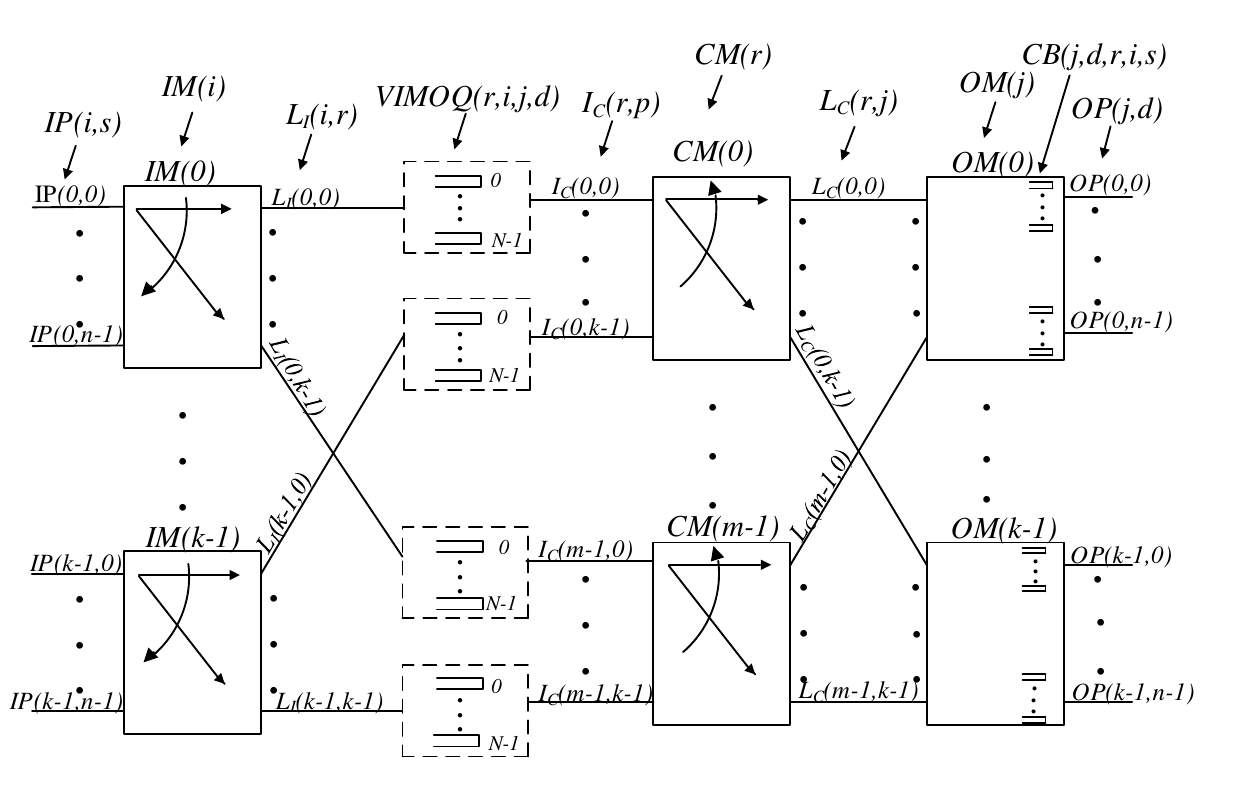}
	\caption{TRIDENT switch.}
	\label{fig:switch}
\end{figure*}

The TRIDENT switch has $N$ inputs and $N$ outputs, each denoted as $IP(i,s)$ and $OP(j,d)$, respectively, where $0 \leq i,~j \leq k-1$, $0 \leq s,~d \leq n-1$, and $N=nk$. Figure \ref{fig:switch} shows the architecture of TRIDENT. This switch has $k$ $n \times m$ IMs, $m$ $k \times k$ CMs, and  $k$ $m \times n$ OMs.
Table \ref{tab:terms} lists the notations used in the description of TRIDENT. 
In the remainder of this paper, we set $n=k=m$ for symmetry and cost-effectiveness. The IMs and CMs are bufferless crossbars while the OMs are buffered ones.
In order to preserve the staggered symmetry and in-order delivery \cite{hu2008joint}, this switch uses a fixed and predetermined configuration sequence, and a reverse desynchronized configuration scheme in CMs. The staggered symmetry and in-order delivery refers to the fact that at time slot $t$, $IP(i,s)$ connects to $CM(r)$ which connects to $OM(j)$. Then at the next time slot $(t+1)$, $IP(i,s)$ connects to $CM((r+1)\mod m)$, which also connects to $OM(j)$. This property enables us to easily represent the configuration of IMs and CMs as a predetermined compound permutation that repeats every $k$ time slots. This property also ensures that cells experience similar delay under uniform traffic, and the incorporation of the in-sequence mechanism enables preserving this delay under nonuniform traffic, as Section \ref{sec:proof} shows.

The switch has virtual input-module output port queues (VIMOQs) between the IMs and CMs to store cells coming from $IM(i)$ and destined to $OP(j,d)$, and each queue is denoted as $VIMOQ(r,i,j,d)$.
Each output of an IM is denoted as $L_{I}(i,r)$. Each output of a VIMOQ is connected to a CM. Each input and output of a CM are denoted as $I_C(r,p)$  and $L_C(r,j)$, respectively. Each OP has $Nk$ crosspoint buffers, each denoted as $CB(r,j,d,i,s)$ and designated for the traffic from each IP traversing different CMs to an OP. A flow control mechanism operates between a CB and VIMOQs to avoid buffer overflow and underflow \cite{cixb}.

Cells are sent from IPs through the IMs for load balancing and then queued at VIMOQs before they are forwarded to their destined OMs through the CMs.
\begin{table}[htbp]
	\caption{Notations used in the description of the TRIDENT switch}
	\vskip .25in
	\label{tab:terms}
	\begin{tabular}{l | p{5cm}}\hline \hline
		Term & Description \\ \hline \hline
		$N$& Number of input/output ports.\\ \hline
		$n$ & Number of input/output ports for each IM and OM.\\ \hline
		$m$ & Number of CMs.\\ \hline
		$k$ & Number of IMs and OMs, where $k =\frac{N}{n}$. \\ \hline
		$IP(i,s)$ &  Input port $s$ of $IM(i)$, where $0 \leq i \leq k-1, 0 \leq s \leq n-1$.\\ \hline
		$IM(i)$ & Input module $i$. \\ \hline
		$CM(r)$ & Central Input Module $r$, where $0 \leq r \leq m-1$.\\ \hline
		$L_{I}(i,r)$ & Output link of $IM(i)$ connected to $CM(r)$. \\ \hline
		$I_{C}(r,p)$ & Input port $p$ of $CM(r)$.\\ \hline
		$L_{C}(r, j)$ & Output link of $CM(r)$ connected to $OM(j)$.\\ \hline
		$VIMOQ(r,i,j,d)$ & VIMOQ at input of CMs that stores cells from $IM(i)$ destined to $OP(j,d)$.\\ \hline
		$CB(r,j,d,i,s)$ & Crosspoint buffer at $OM(j)$ that stores cells from $IP(i,s)$ going through $CM(r)$ and destined to $OP(j,d)$. \\ \hline
		$OP(j,d)$ & Output port $d$ at $OM(j)$. \\ \hline
	\end{tabular}
\end{table}

\subsection{Module Configuration}
\label{sec:operation-of-lbc}
The IMs are configured based on a predetermined sequence of $k$ disjoint permutations, where one permutation is applied each time slot.  We call a permutation {\it disjoint} from the set of permutations if the input-output pair interconnection is unique in one and only one of the $k$ permutations. Cells at the inputs of IMs are forwarded to the outputs of the IMs determined by the configuration at that time slot. A cell is then stored in the VIMOQ corresponding to its destination OP. 

Similar to the IMs, CMs are configured based on a predetermined sequence of $k$ disjoint permutations. Unlike IMs, CMs follow a desynchronized configuration; a different permutation is used each time slots, and the configuration follows a cycle but in counter clock manner to that of the IM. The Head-of-Line (HoL) cell at the VIMOQ destined to $OP(j,d)$ is forwarded to its destination when the input of the CM is connected to the input of the destined $OM(j)$. Else, the HoL cell waits until the required configuration takes place. The forwarded cell is queued at the CB of its destination OP once it arrives in the OM. 

The configurations of the bufferless IMs and CMs are as follows. At time slot $t$, IM input $IP(i,s)$ is interconnected to IM output $L_{I}(i,r)$, as follows:
\begin{equation}
\label{equ1}
r = (s + t) \mod m
\end{equation}
and each CM input $I_{C}(r,p)$ is interconnected to output $L_{C}(r, j)$ as follows:
\begin{equation}
\label{eq:LII}
j  = (p - t + r) \mod k. 
\end{equation}
The use of CBs at an OP allows forwarding a cell from of a VIMOQ to its destined output without requiring port matching \cite{lin2013minimizing}.

\begin{figure}[htbp]
	\centering
	\subfigure[Time slot 0]{
		\includegraphics[width=0.9\columnwidth]{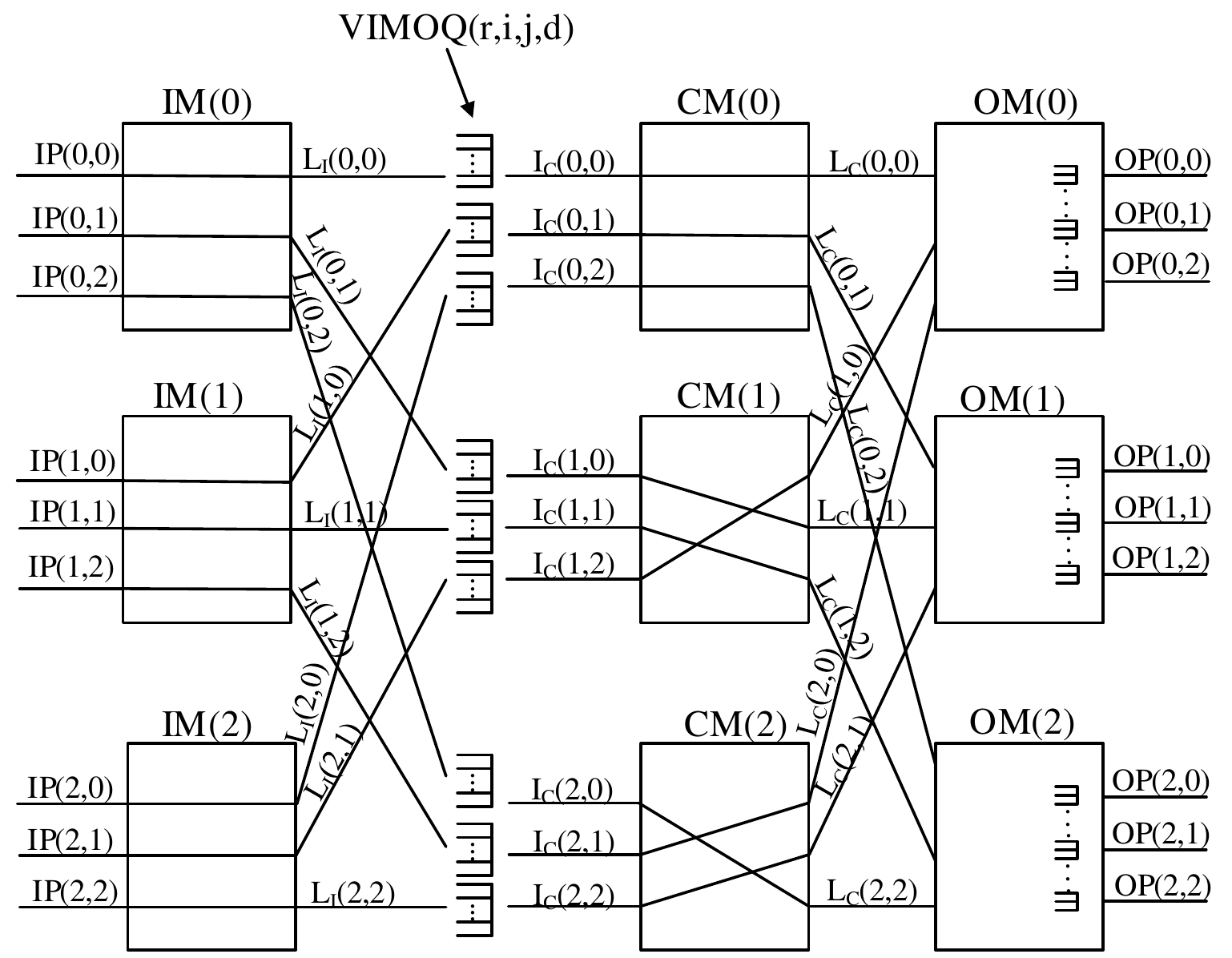}
		\label{fig:SwitchConfig1}}
	\subfigure[Time slot 1]{
		\includegraphics[width=0.9\columnwidth]{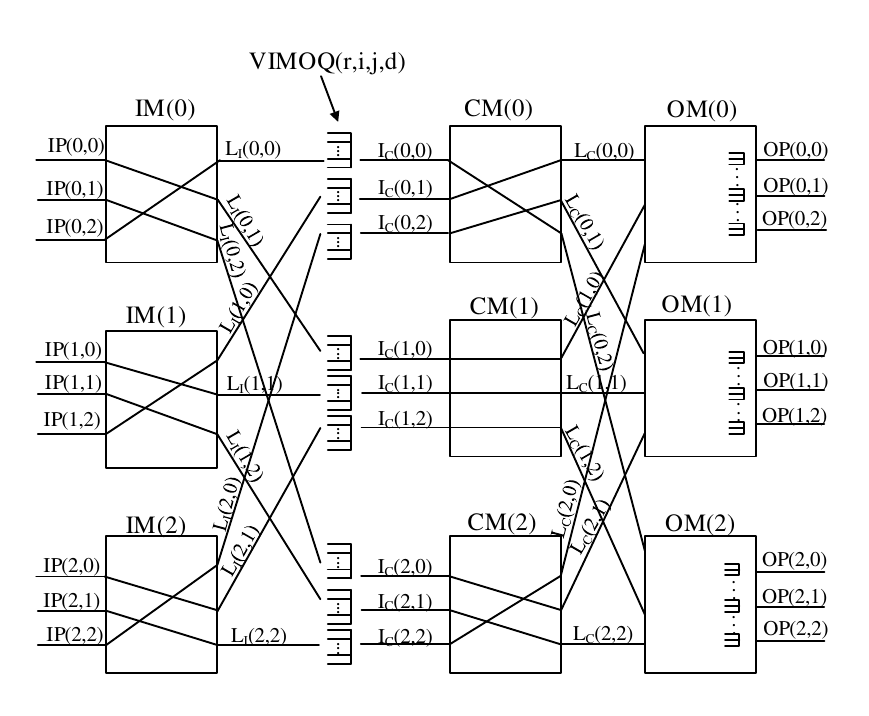}
		\label{fig:SwitchConfig2}}
	\subfigure[Time slot 2]{
		\includegraphics[width=0.9\columnwidth]{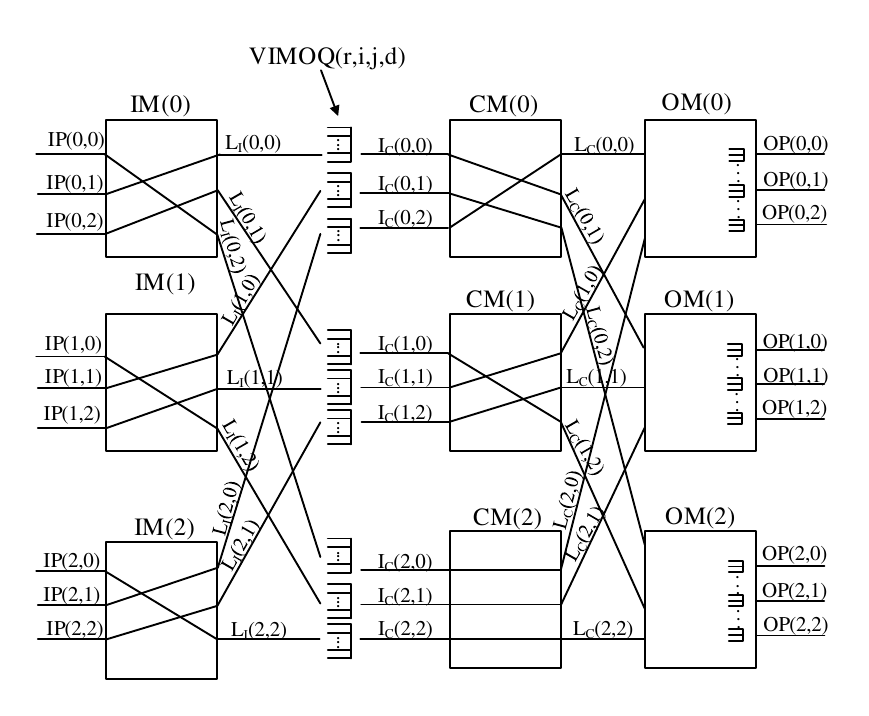}
		\label{fig:SwitchConfig3}}
	\caption{Configuration example of a 9 $\times$ 9 TRIDENT switch modules.}
	\label{fig:switchconfig}
\end{figure}

Table \ref{table:configuration} shows an example of the configuration of the IMs and CMs of a 9$\times$9 TRIDENT switch. Because $k=3$, the example shows the configuration of three consecutive time slots. In this table, we use $ w \rightarrow x$ to denote an interconnection between $w$ and $x$.
Figure \ref{fig:switchconfig} shows the configuration of the modules. 

\begin{table*}[htbp]
	\centering
	\caption{Example of configuration of modules in a 9 $\times$ 9 TRIDENT switch.}
	\vskip .25in
	\label{table:configuration}
	\begin{tabular}{ |c|c|c|c|c|c|c| }
		\hline
		\multicolumn{7}{ |c| }{Configuration} \\
		\hline
		Time slot & $IM(0)$ & $CM(0)$& $IM(1)$ & $CM(1)$ & $IM(2)$ & $CM(2)$ \\ \hline
		\multirow{3}{*}{$t=0$}
		& $IP(0,0) \rightarrow L_{I}(0,0)$ & $I_c(0,0) \rightarrow L_{C}(0,0)$ & $IP(1,0) \rightarrow L_{I}(1,0)$ & $I_c(1,0) \rightarrow L_{C}(1,1)$& $IP(2,0) \rightarrow L_{I}(2,0)$ & $I_c(2,0) \rightarrow L_{C}(2,2)$\\
		& $IP(0,1) \rightarrow L_{I}(0,1)$ & $I_c(0,1) \rightarrow L_{C}(0,1)$ 	& $IP(1,1) \rightarrow L_{I}(1,1)$ & $I_c(1,1) \rightarrow L_{C}(1,2)$ & $IP(2,1) \rightarrow L_{I}(2,1)$ & $I_c(2,1) \rightarrow L_{C}(2,0)$\\
		& $IP(0,2) \rightarrow L_{I}(0,2)$ & $I_c(0,2) \rightarrow L_{C}(0,2)$& $IP(1,2) \rightarrow L_{I}(1,2)$ & $I_c(1,2) \rightarrow L_{C}(1,0)$ & $IP(2,2) \rightarrow L_{I}(2,2)$ & $I_c(2,2) \rightarrow L_{C}(2,1)$\\ \hline
		\multirow{3}{*}{$t=1$} 
		& $IP(0,0) \rightarrow L_{I}(0,1)$ & $I_c(0,0) \rightarrow L_{C}(0,2)$& $IP(1,0) \rightarrow L_{I}(1,1)$ & $I_c(1,0) \rightarrow L_{C}(1,0)$& $IP(2,0) \rightarrow L_{I}(2,1)$ & $I_c(2,0) \rightarrow L_{C}(2,1)$\\
		& $IP(0,1) \rightarrow L_{I}(0,2)$ & $I_c(0,1) \rightarrow L_{C}(0,0)$& $IP(1,1) \rightarrow L_{I}(1,2)$ & $I_c(1,1) \rightarrow L_{C}(1,1)$ & $IP(2,1) \rightarrow L_{I}(2,2)$ & $I_c(2,1) \rightarrow L_{C}(2,2)$\\
		& $IP(0,2) \rightarrow L_{I}(0,0)$ & $I_c(0,2) \rightarrow L_{C}(0,1)$&$IP(1,2) \rightarrow L_{I}(1,0)$ & $I_c(1,2) \rightarrow L_{C}(1,2)$&$IP(2,2) \rightarrow L_{I}(2,0)$ & $I_c(2,2) \rightarrow L_{C}(2,0)$\\ 
		\hline
		\multirow{3}{*}{$t=2$} 
		& $IP(0,0) \rightarrow L_{I}(0,2)$ & $I_c(0,0) \rightarrow L_{C}(0,1)$	& $IP(1,0) \rightarrow L_{I}(1,2)$ & $I_c(1,0) \rightarrow L_{C}(1,2)$& $IP(2,0) \rightarrow L_{I}(2,2)$ & $I_c(2,0) \rightarrow L_{C}(2,0)$\\
		& $IP(0,1) \rightarrow L_{I}(0,0)$ & $I_c(0,1) \rightarrow L_{C}(0,2)$	& $IP(1,1) \rightarrow L_{I}(1,0)$ & $I_c(1,1) \rightarrow L_{C}(1,0)$	& $IP(2,1) \rightarrow L_{I}(2,0)$ & $I_c(2,1) \rightarrow L_{C}(2,1)$\\
		& $IP(0,2) \rightarrow L_{I}(0,1)$ &  $I_c(0,2) \rightarrow L_{C}(0,0)$& $IP(1,2) \rightarrow L_{I}(1,1)$ & $I_c(1,2) \rightarrow L_{C}(1,1)$& $IP(2,2) \rightarrow L_{I}(2,1)$ & $I_c(2,2) \rightarrow L_{C}(2,2)$\\ 
		\hline
		\hline
		
	\end{tabular}
\end{table*}

\subsection{Arbitration at Output Ports}
\label{sec:OP-arbitration}
Each output port has a round-robin arbiter to keep track of the next flow to serve, and $N$ flow pointers to keep track of the next cell to serve for each flow. Here, a flow is the set of cells from $IP(i,s)$ destined to $OP(j,d)$. An output port arbiter selects the flow to serve in a round-robin fashion. For this selection, the output arbiter selects the HoL cell of a CB if the cell's order matches the expected cell order for that flow.
Because the output port arbiter selects the older cell based on the order of arrival to the switch, this selection prevents out-of-sequence forwarding. We discuss this property in Section \ref{sec:proof}. Furthermore, the round-robin schedule ensures fair service for different flows. If there is no HoL cell with the expected value for a particular flow, the arbiter moves to the next flow. 

\subsection{Analysis of Crosspoint Buffer Size }
\label{proof-cb}
In this section, we show that no CB queue in the switch receives more than one cell in a time slot and those who receive cells at a rate of $1/kN$ are served at rates of $1/kN$.
Let us consider a scenario where all the IPs in the switch  only have traffic for one OP. 
The largest admissible arrival rate at an IP is:
\begin{equation}
\label{IP_analysis_1}
\lambda_{i,s,j,d} = \frac{1}{N}
\end{equation}
The input load, $\lambda_{i,s,j,d}$, gets load-balanced to VIMOQs at a rate of $\frac{1}{m}$. The aggregate traffic arrival rate at a VIMOQ from an IM, $R_{V}$, is:
\begin{equation}
\label{CM_analysis_1}
{R_{V}} =  \frac{1}{m}  \displaystyle\sum_{i=0}^{n} \lambda_{i,s,j,d} = \frac{n}{mN}
\end{equation}
because $m=n=k$, therefore,
\begin{equation}
\label{CM_analysis_2}
{R_{V}}  = \frac{1}{N}
\end{equation}
The aggregate traffic rate at a CM for an OP is:
\begin{equation}
\label{CM_analysis_3}
{R_{CM}}  = \displaystyle\sum_{}^{k} \frac{1}{N} =  \frac{1}{k}
\end{equation}
The traffic arrival rate to a CB, $R_{C}$, is the aggregate traffic from an IP through a CM or:
\begin{equation}
\label{CB_analysis_1}
R_{C} =   \frac{1}{N} {R_{CM}} = \frac{1}{kN} 
\end{equation}
Therefore, ${R_{C}} \leq {S_{C}}$ for admissible traffic, which implies that the crosspoint buffer size at OMs does not impact the performance of the switch because the queue size does not grow with the input load.

\subsection{In-sequence Cell Forwarding Mechanism}
\label{sec:in-sequence-forwarding} 
The proposed in-sequence forwarding mechanism of TRIDENT is based on tagging cells of a flow at the inputs with their arriving sequence number, and forwarding cells from the crosspoint buffers to the output port in the same sequence they arrived in the input. The policy used for keeping cells in-sequence is as follows: 
When a cell of a flow arrives in the input port, the input port arbiter appends the arrival order to the cell (for the corresponding flow).
After being forwarded through $L_I(i,r)$, the cell is stored at the VIMOQ for the destination OP. When the CM configuration permits, the cell is forwarded to the destined OM and stored at the queue for traffic from the IP to the destined OP traversing that CM.
An OP arbiter selects cells of a flow in the order they arrived in the switch by using the arrival order carried by each cell. As an example of this operation, Table \ref{table:cell-arrival} shows the arrival times of cell $c_{1,1}$, $c_{2,1}$, and $c_{2,2}$, where $c_{y, t_x}$ denotes flow $y$ and arrival time $t_x$ to the VIMOQs. Cell $c_{2,1}$ is queued behind $c_{1,1}$, and $c_{2,2}$ is placed in an empty VIMOQ. Table \ref{table:2stage-departures-from-voqcm-example} shows the time slots when the cells are forwarded from the VIMOQ. For example, when $c_{2,2}$ leaves the VIMOQ before $c_{2,1}$. Table \ref{table:2stage-departures-from-cbs-example} shows the time slots when the cells are forwarded to the destination OP after the output-port arbitration is performed.

\begin{table}[htbp]
	\centering
	\caption{Time slots of cell arrival to VIMOQs in example of the in-sequence forwarding mechanism.}
	\vskip .25in
	\label{table:cell-arrival}
		\begin{tabular}{ | c | c | c |  } 
			\hline
			\multicolumn{3}{|c|}{Cell arrival time} \\ \hline
			$t_{x}$ & $t_{x+1}$ & $t_{x+2}$ \\ 
			\hline
			$c_{1,1}$ && \\ 
			\hline
			& $c_{2,1}$ & $c_{2,2}$  \\ 
			\hline
		\end{tabular}
\end{table}

\begin{table}[htbp]
	\centering
	\caption{Time slots of cells departure from VIMOQs in example of the in-sequence forwarding mechanism.}
	\vskip .25in
	\label{table:2stage-departures-from-voqcm-example}
		\begin{tabular}{ | c | c | c | c | c | c | c |}
			\hline
			\multicolumn{7}{|c|}{Cell departure time slots from VIMOQs} \\
			\hline
			$t_{x}$ & $t_{x+1}$ & $t_{x+2}$ &$t_{x+3}$ & $t_{x+4}$ & $t_{x+5}$ &$t_{x+6}$  \\ 
			\hline
			&&& $c_{1,1}$ &&&\\ 
			\hline
			&&&& $c_{2,2}$ && $c_{2,1}$ \\
			\hline
		\end{tabular}
\end{table}

\begin{table}[htbp]
	\centering
	\caption{Time slots of cells departure from CBs in example of the in-sequence forwarding mechanism.}
	\vskip .25in
	\label{table:2stage-departures-from-cbs-example}
		\begin{tabular}{ | c | c | c | c | c | c | c | c | c |}
			\hline
			\multicolumn{9}{|c|}{Cell departure time slots from CBs} \\
			\hline
			$t_{x}$ & $t_{x+1}$ & $t_{x+2}$ &$t_{x+3}$ & $t_{x+4}$ & $t_{x+5}$ &$t_{x+6}$ &$t_{x+7}$ &$t_{x+8}$  \\ 
			\hline
			&&&& $c_{1,1}$ &&&&\\ 
			\hline
			&&&&&&&$c_{2,1}$&$c_{2,2}$\\ 
			\hline
		\end{tabular}
\end{table}

\begin{figure}[htbp]
	\includegraphics[width=\columnwidth]{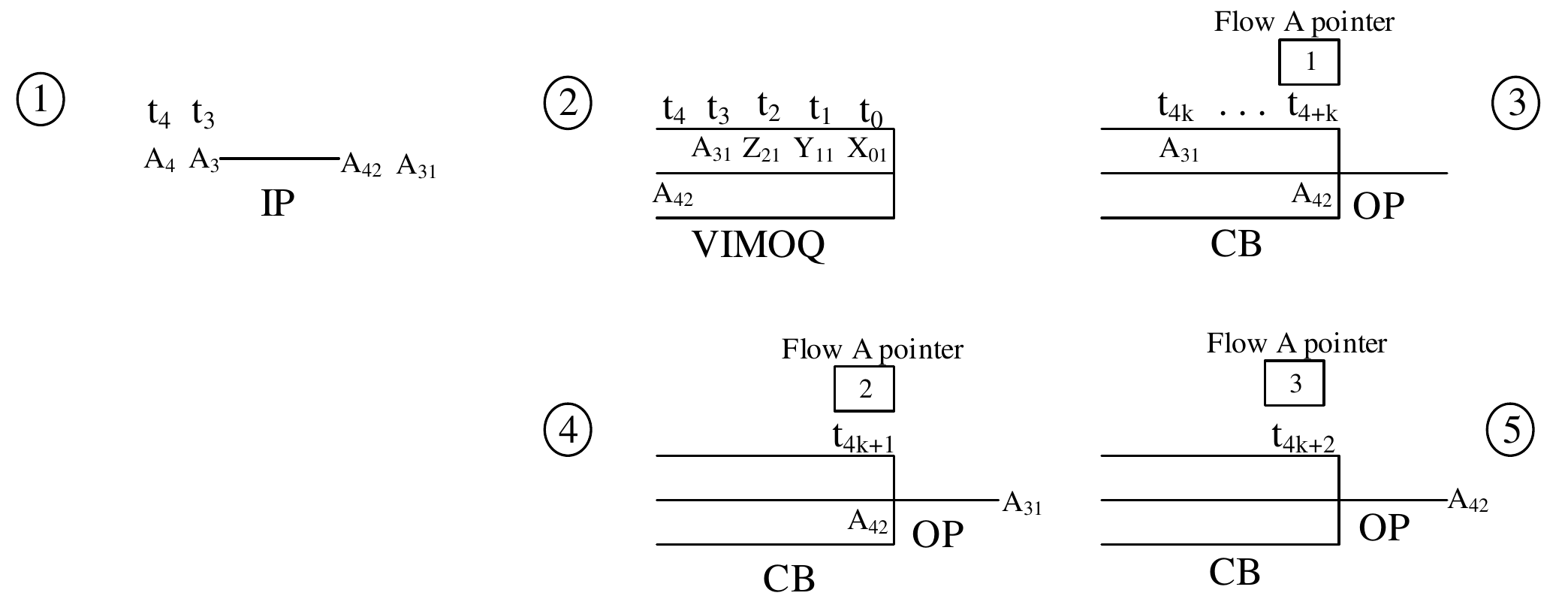}
	\caption{Example of TRIDENT In-sequence Mechanism.}
	\label{ex:trident-in-sequence}
\end{figure}

Figure \ref{ex:trident-in-sequence} shows a single flow $A$ with two cells, $A_3$ and $A_4$, arriving at timeslots, $t_3$ and $t_4$, respectively. Let us assume that no cell of this flow has transited the switch. The cell that arrives at $t_3$ is appended a tag of $1$ (i.e., the order of arrival) and the cell that arrives at $t_4$ is appended a tag of $2$. Both cells are load balanced and forwarded to different virtual input module output queues (VIMOQs). As shown in Step 2 of Figure \ref{ex:trident-in-sequence}, $A_{31}$ is forwarded to a queue with cells from other flows, while $A_{42}$, the younger cell, is forwarded to an empty queue.
Therefore, $A_{42}$ arrives at the output port (OP) before $A_{31}$ (Step 3). Because the pointer of flow $A$ at this OP has not received any cell for this flow, it currently points to tag $1$. Hence $A_{42}$ remains at the CB until $A_{31}$ arrives and is forwarded out the OP. Thereafter, flow $A$ pointer at this OP is updated to $2$ and $A_{42}$ is forwarded out the OP.

\section{Throughput Analysis}
\label{sec:throughput-analysis}
In this section, we analyze the performance of the proposed TRIDENT switch. 


Let us denote the traffic coming to the IMs, CMs, OMs, OPs, and the traffic leaving TRIDENT as $\mathbf{R_1}$, $\mathbf{R_2}$, $\mathbf{R_3}$, $\mathbf{R_4}$ and ${R_5}$, respectively. Here, $\mathbf{R_1}$ and $\mathbf{R_2}$, and $\mathbf{R_3}$ are $N \times N$ matrices, $\mathbf{R_4}$ comprises $N$ $N \times 1$ column vectors, and $R_5$ comprises $N$ scalars. Figure~\ref{fig:switch} shows these traffic points set at each stage of TRIDENT with the corresponding labels at the bottom of the figure.

The traffic from input ports to the IM stage, $\mathbf{R_1}$, is defined as:
\begin{equation}
\label{eq:R1}
\mathbf{R_1}=[\lambda_{u,v}]
\end{equation}
where, $\lambda_{u,v}$ is the arrival rate of traffic from input $u$ to output $v$, and
\begin{equation}
\label{equ:u}
u=ik+s
\end{equation}
\begin{equation}
\label{equ:v}
v=jm+d
\end{equation} 
where $0 \leq u, v \leq N-1$. 

In the following analysis, we consider admissible traffic, which is defined as:
\begin{equation}
\label{eq:admissible}
\sum_{u=0} ^ {N-1} \lambda_{u,v} \leq 1, \  \sum_{v=0} ^ {N-1} \lambda_{u,v} \leq 1
\end{equation}
and as i.i.d. traffic.

The IM stage of TRIDENT balances the traffic load coming from the input ports to the VIMOQs. Specifically, the permutations used to configure the IMs forwards the traffic from an input to $k$ different CMs, and then to the VIMOQs connected to these CMs in $k$ consecutive time slots.

$\mathbf{R_2}$ is the traffic directed towards CMs and it is derived from $\mathbf{R_1}$ and the permutations of IMs.
The configuration of the IM stage at time slot $t$ that connects $IP(i,s)$ to $L_{I}(i,r)$ are represented as an $N \times N$ permutation matrix, $\mathbf{\Pi}(t)=[\pi_{u, v}]$, where $r$ is determined from (\ref{equ1}) and the matrix element:\\
$\pi_{u,\upsilon} = \begin{dcases*}
1  & for any $u$, $\upsilon= rk+i $\\
0 & elsewhere. 
\end{dcases*}
\label{eq:4a}$

The configuration of the IM stage can be represented as a compound permutation matrix, 
$\mathbf{P_{1}}$, which is the sum of the IM permutations over $k$ time slots as follows, \\
$\mathbf{P_1} = \displaystyle\sum_{}^{k} \mathbf{\Pi}(t)
\label{eq:P1}$

Because the configuration is repeated every $k$ time slots, the traffic load from the same input going to each VIMOQ is $\frac{1}{k}$ of the traffic load of $\mathbf{R_1}$. 
Therefore, a row of $\mathbf{R_2}$ is the sum of the row elements of $\mathbf{R_1}$ at the non zero positions of $\mathbf{P_1}$, normalized by $k$.
This is:
\begin{equation}\label{eq:R2}
\mathbf{R_2} = \frac{1}{k}((\mathbf{R_1}*\mathbb{1}) \circ \mathbf{P_1})
\end{equation}
where $\mathbb{1}$ denotes an $N \times N$ unit matrix and $\circ$ denotes element/position wise multiplication.
There are $k$ non-zero elements in each row of $\mathbf{R_2}$. Here, $\mathbf{R_2}$ is the aggregate traffic in all the VIMOQs destined to all OPs. This matrix can be further decomposed into $k$ $N \times N$ submatrices, $\mathbf{R_2}(j,d)$, each of which is the aggregate traffic at VIMOQs designated for $OP(j,d)$.
\begin{equation}\label{equ:R2_2}
\mathbf{R_2} = \displaystyle\sum_{}^{k} \displaystyle\sum_{d=0}^{k-1} \mathbf{R_2}(j,d)
\end{equation} 
where $j$ is obtained from (\ref{equ:v}) $\forall \;d$ and $d$ is also obtained from (\ref{equ:v}) but for the different $j$.
The configuration of the CM stage at time slot $t$ that connects $I_{c}(r,p)$ to $L_{C(r,j)}$ 
may be represented as an $N \times N$ permutation matrix, 
$\mathbf{\Phi}(t)=[\phi_{u,v}]$, where $j$ is determined from (\ref{eq:LII}) and the matrix element:\\
$\phi_{u,v} = \begin{dcases*}
1  & for any $u$,~$v= jk+r $\\ 
0 & elsewhere. 
\end{dcases*}
\label{eq:4}$

Similarly, the switching process at the CM stage is represented by a compound permutation matrix $\mathbf{P_2}$, which is the sum of $k$ permutations used at the CM stage over $k$ time slots. 
Here,\\
$\mathbf{P_2} = \displaystyle\sum_{t=0}^{k-1}\mathbf{\Phi}(t)
\label{eq:P2}$\\
The traffic destined to $OP(j,d)$ at $OM(j)$, $\mathbf{R_3}(j,d)$, is:
\begin{equation}\label{R3}
\mathbf{R_3}(j,d) = \mathbf{R_2}(j,d) \circ \mathbf{P_2}
\end{equation}
The aggregate traffic at CBs of an OP for the different IPs, $\mathbf{R_4}(v)$, is obtained from the multiplication of $\mathbf{R_{3}}(j,d)$ with a vector of all ones, $\vec{1}$, or:
\begin{equation}\label{equ9}
\mathbf{R_4}(v) = \mathbf{R_{3}}(j,d) * \vec{1}
\end{equation}
Each row of $\mathbf{R_4}(v)$ is the aggregate traffic at the CBs from each IP.
The traffic leaving an OP, ${R_5}(v)$, is:
\begin{equation}\label{eq:R5}
{R_5}(v) = (\vec{1})^T * \mathbf{R_4}(v) 
\end{equation}
Therefore, ${R_5}(v)$ is the sum of the traffic leaving $OP(v)$.

The following example shows the operations performed on traffic coming to a 4$\times$4 ($k= 2$) TRIDENT switch. Let the input traffic matrix be
\[
\  \mathbf{R_1} =
\begin{bmatrix}
\setlength{\arraycolsep}{3pt}
\lambda_{0,0}  & \lambda_{0,1} & \lambda_{0,2} & \lambda_{0,3} \\
\lambda_{1,0}  & \lambda_{1,1} & \lambda_{1,2} & \lambda_{1,3} \\
\lambda_{2,0}  & \lambda_{2,1} & \lambda_{2,2} & \lambda_{2,3} \\
\lambda_{3,0}  & \lambda_{3,1} & \lambda_{3,2} & \lambda_{3,3} \\
\end{bmatrix}
\]
Then, $\mathbf{R_2}$ is generated from the arriving traffic and the configuration of IM.
The compound permutation matrix for the IM stage for this switch is:
\[
\mathbf{P_1} =
\begin{bmatrix}
\setlength{\arraycolsep}{3pt}
1  & 0  & 1 & 0\\
1  & 0  & 1 & 0\\
0  & 1  & 0 & 1 \\
0  & 1  & 0 & 1
\end{bmatrix}
\]
Using (\ref{eq:R2}), we get
\[
\mathbf{R_2} = 1/2
\begin{bmatrix}
\setlength{\arraycolsep}{3pt}
\sum_{i=0}^ {3} \lambda_{0i}  & 0 & \sum_{i=0}^ {3} \lambda_{0i} & 0 \\
\sum_{i=0}^ {3} \lambda_{1i}  & 0 & \sum_{i=0}^ {3} \lambda_{1i} & 0 \\
0  & \sum_{i=0}^ {3} \lambda_{2i} & 0 & \sum_{i=0}^ {3} \lambda_{2i} \\
0  & \sum_{i=0}^ {3} \lambda_{3i} & 0 & \sum_{i=0}^ {3} \lambda_{3i} \\
\end{bmatrix}
\]
From (\ref{equ:R2_2}), the traffic matrix at VIMOQs destined for the different OMs are:\\
\[
\mathbf{R_2}(0) = \frac{1}{2}
\begin{bmatrix}
\setlength{\arraycolsep}{3pt}
\lambda_{0,0}+\lambda_{0,1}  & 0 & \lambda_{0,0}+\lambda_{0,1} & 0 \\
\lambda_{1,0} + \lambda_{1,1} & 0 & \lambda_{1,0} + \lambda_{1,1}   & 0 \\
0  & \lambda_{2,0} + \lambda_{2,1}   & 0 & \lambda_{2,0} + \lambda_{2,1}    \\
0 & \lambda_{3,0} + \lambda_{3,1}    & 0  & \lambda_{3,0} + \lambda_{3,1}  \\
\end{bmatrix}
\]
\[
\mathbf{R_2}(1) = \frac{1}{2}
\begin{bmatrix}
\setlength{\arraycolsep}{3pt}
\lambda_{0,2}+\lambda_{0,3}  & 0 & \lambda_{0,2}+\lambda_{0,3} & 0 \\
\lambda_{1,2} + \lambda_{1,3} & 0 & \lambda_{1,2} + \lambda_{1,3}   & 0 \\
0  & \lambda_{2,2} + \lambda_{2,3}   & 0 & \lambda_{2,2} + \lambda_{2,3}    \\
0 & \lambda_{3,2} + \lambda_{3,3}    & 0  & \lambda_{3,2} + \lambda_{3,3}  \\
\end{bmatrix}
\]
The rows of $\mathbf{R_2}(v)$ represent the traffic from IPs, and the columns represent $VIMOQ(r,i,j,d)$ at $I_C(r,p)$.
The compound permutation matrix for the CM stage for this switch is:
\[
\mathbf{P_2} =
\begin{bmatrix}
\setlength{\arraycolsep}{3pt}
1  & 0  & 1 & 0\\
1  & 0  & 1 & 0 \\
0  & 1  & 0 & 1 \\
0  & 1  & 0 & 1
\end{bmatrix}
\]
From (\ref{R3}),  the traffic forwarded to an OP is:\\
$
\mathbf{R_3}(0,0) = \frac{1}{2}
\begin{bmatrix}
\setlength{\arraycolsep}{3pt}
\lambda_{0,0}   & 0 & \lambda_{0,0}  & 0 \\
\lambda_{1,0}  & 0 & \lambda_{1,0}  & 0 \\
0  & \lambda_{2,0}  & 0 & \lambda_{2,0} \\
0 & \lambda_{3,0}   & 0  & \lambda_{3,0} \\
\end{bmatrix}
\\
\mathbf{R_3}(0,1) = \frac{1}{2}
\begin{bmatrix}
\setlength{\arraycolsep}{3pt}
\lambda_{0,1}  & 0 & \lambda_{0,1}  & 0 \\
\lambda_{1,1}  & 0 & \lambda_{1,1} & 0 \\
0  & \lambda_{2,1}  & \lambda_{2,1} \\
0 & \lambda_{3,1}  & 0  & \lambda_{3,1}  \\
\end{bmatrix}
\\
\mathbf{R_3}(1,0) = \frac{1}{2}
\begin{bmatrix}
\setlength{\arraycolsep}{3pt}
\lambda_{0,2}  & 0 & \lambda_{0,2}  & 0 \\
\lambda_{1,2}  & 0 & \lambda_{1,2}  & 0 \\
0  & \lambda_{2,2}  & 0 & \lambda_{2,2}  \\
0 & \lambda_{3,2}   & 0  & \lambda_{3,2}  \\
\end{bmatrix}
\\
\mathbf{R_3}(1,1) = \frac{1}{2}
\begin{bmatrix}
\setlength{\arraycolsep}{3pt}
\lambda_{0,3}  & 0 & \lambda_{0,3} & 0 \\
\lambda_{1,3}  & 0 &  \lambda_{1,3}  & 0 \\
0  & \lambda_{2,3}  & 0 & \lambda_{2,3}  \\
0 & \lambda_{3,3}   & 0  &  \lambda_{3,3} \\
\end{bmatrix}
$\\
The rows of $\mathbf{R_3}(j,d)$ represent the traffic from $VIMOQ(r,i,j)$ at $I_C(r,p)$ and the columns represent $L_C(r,j)$.\\
The traffic forwarded from $CB$s allocated for the different IPs to the corresponding OP is obtained from 
(\ref{equ9}): 
$\mathbf{R_4}(0) = 
\begin{bmatrix}
\setlength{\arraycolsep}{3pt}
\lambda_{0,0} \\
\lambda_{1,0} \\
\lambda_{2,0} \\
\lambda_{3,0}\\
\end{bmatrix}
$,\qquad$
\mathbf{R_4}(1) = 
\begin{bmatrix}
\setlength{\arraycolsep}{3pt}
\lambda_{0,1}  \\
\lambda_{1,1} \\
\lambda_{2,1}\\
\lambda_{3,1}\\
\end{bmatrix}
\\
\mathbf{R_4}(2)= 
\begin{bmatrix}
\setlength{\arraycolsep}{3pt}
\lambda_{0,2}   \\
\lambda_{1,2} \\
\lambda_{2,2}  \\
\lambda_{3,2}  \\
\end{bmatrix}
$,\qquad$
\mathbf{R_4}(3)= 
\begin{bmatrix}
\setlength{\arraycolsep}{3pt}
\lambda_{0,3} \\
\lambda_{1,3} \\
\lambda_{2,3} \\
\lambda_{3,3}  \\
\end{bmatrix}
\label{ex:R_4}
$\\
The rows of $\mathbf{R_4}(v)$ represent the traffic from $IP(i,s)$.
Using (\ref{eq:R5}), we obtain the sum of the traffic leaving the OP, or:\\
$
{R_5}(0)=
\sum_{i=0}^ {3} \lambda_{i0}
$,\qquad $
{R_5}(1)=
\sum_{i=0}^ {3} \lambda_{i1}
$,\qquad$
{R_5}(2)=
\sum_{i=0}^ {3} \lambda_{i2}
$,\qquad$
{R_5}(3)=
\sum_{i=0}^ {3} \lambda_{i3}
$

As raised from the example, one may wonder if TRIDENT achieves 100\% throughput. This property of TRIDENT is discussed as follows:

From $\mathbf{R_4}(0)$ to $\mathbf{R_4}(3)$ above, we can deduce that $\mathbf{R_4}$ is equal to the input traffic $\mathbf{R_1}$, or, in general:
\begin{equation}
\label{theorem1:eq1}
\mathbf{R_4}(v) = \mathbf{R_1}(v) \; \forall \; v
\end{equation} 
Also, because $\mathbf{R_2}$ and $\mathbf{R_4}(v)$ meet the admissibility condition in (\ref{eq:admissible}), and ${R_5}(v)$ does not exceed the traffic rate for any $OP(v)$, the aggregated traffic loads at each VIMOQ, CB, and OP do not exceed the capacity of each output link.
From the admissibility of $\mathbf{R_2}$ and $\mathbf{R_4}(v)$, and (\ref{theorem1:eq1}), we can infer that the input traffic is fully forwarded to the output ports.

As discussed in Section \ref{sec:OP-arbitration}, an output arbiter selects a flow in a round-robin fashion and a cell of that flow based on the arrival order. If a cell of a flow is not selected, the OP arbiter moves to the next flow. This arbitration scheme ensures fairness and that the cells forwarded to the OP are also forwarded out of the OP. Hence, from $R_5(0)$ to $R_5(3)$, we can infer that ${R_5}(v)$ is equal to $\mathbf{R_4}(v)$, or:
\begin{equation}
\label{theorem1:eq2}
{R_5}(v) = (\vec{1})^T * \mathbf{R_4}(v) \; \forall \; v
\end{equation} 
From (\ref{theorem1:eq1}) and (\ref{theorem1:eq2}), we can conclude that TRIDENT achieves 100\% throughput under admissible i.i.d. traffic. We present the proof in Section \ref{tri:stability}.
\\

\section{100\% Throughput}
\label{tri:stability}
	In this section we prove that TRIDENT  achieves 100\% throughput by using the analysis under admissible i.i.d traffic.
	\label{TRIDENT:stability}
	\begin{theorem}
		\label{theorem:throughput_tri}
		{\it TRIDENT achieves 100\% throughput under admissible i.i.d traffic.}
	\end{theorem}
	\noindent {\bf Proof:}
	Here, we proof that TRIDENT achieves 100\% throughput. This is achieved by showing that VIMOQs and CBs are weakly stable under i.i.d. traffic. Because a stable switch achieves 100\% throughput under admissible i.i.d traffic \cite{mekkittikul1998practical}. A switch is considered stable under a traffic distribution if the queue length is bounded.
	The queues are considered to be weakly stable if the queue occupancy drift from its initial state is finite $\epsilon$ ~ $\forall ~ t$ as $\lim_{t\to\infty}$.
		Let us represent the queue occupancy of VIMOQs at time slot $t$, $\mathbf{N_\mu}(t)$ as:
		\begin{equation}
		\label{N_{u-1_tri}}
		\mathbf{N_\mu}(t) = \mathbf{N_\mu}(t-1) + \mathbf{A_\mu}(t) - \mathbf{D_\mu}(t)
		\end{equation}
		where $\mathbf{A_\mu}(t)$ is the aggregate traffic arrival matrix at time slot $t$ to VIMOQs and $\mathbf{D_\mu}(t)$ is the service rate matrix of VIMOQs at time slot $t$.  
		Solving (\ref{N_{u-1_tri}}) with an initial condition $\mathbf{N_\mu}(0)$, recursively yields:
		\begin{equation}
		\label{N2}
		\mathbf{N_\mu}(t) = \mathbf{N_\mu}(0) + \displaystyle\sum_{\gamma=0}^{t} \mathbf{A_\mu}(\gamma) - \displaystyle\sum_{\gamma=0}^{t} D_\mu(\gamma)
		\end{equation}
		Because a VIMOQ is serviced at least once every $N$ time slots, the service rate of a VIMOQ at a CM for $OP(v)$ at time slot $t$, $d_{{\mu}_{v}}(t)$ is:
		\[d_{{\mu}_{v}}(t)= \frac{1}{N}~ \forall ~ \mu ~ and ~ v\]
		Then, the service matrix of VIMOQs is:
		\begin{equation}
		\label{eq:D2_tri}
		\mathbf{D_{\mu}}(t)=[d_{{\mu}_{v}}(t)]
		\end{equation}
		and representing $\mathbf{R_2}$ as the aggregate traffic arrival to VIMOQs or:
		\begin{equation}
		\label{tri-vomq-matrix-2a}
		\mathbf{R_2} = \displaystyle\sum_{\gamma=0}^{t} \mathbf{A_2}(\gamma) 
		\end{equation} 
		Substituting (\ref{eq:D2_tri}) and (\ref{tri-vomq-matrix-2a}) into (\ref{N2}) gives:
		\begin{equation}
		\label{tri-vomq-matrix-4}
		\mathbf{N_{\mu}}(t) = \mathbf{N_\mu}(0) +  \mathbf{R_2} - \frac{1}{N} \mathbf{P_1}
		\end{equation}
		\begin{equation}
		\mathbf{R_2} - \frac{1}{N} \mathbf{P_1} \leq \epsilon < \infty
		\label{tri-vomq-matrix-5}
		\end{equation}
		We recall from section III.A that $\mathbf{R_2}$ is admissible, and by substituting $\mathbf{P_1}$ and $\mathbf{R_2}$ into (\ref{tri-vomq-matrix-5}), shows that $\epsilon$ is finite.
		We can conclude from (\ref{tri-vomq-matrix-4}) and (\ref{tri-vomq-matrix-5}), that the occupancy of VIMOQ is weakly stable.
		$\blacksquare$ \\
		We now prove the stability of CBs. The queue occupancy matrix of CBs at time slot $t$ can be represented as:
		\begin{equation}
		\label{tri-cb-matrix-1}
		\mathbf{N_c}(t) = \mathbf{N_c}(t-1) + \mathbf{A_c}(t) - \mathbf{D_c}(t)
		\end{equation}
		where $\mathbf{A_c}(t)$ is the aggregate traffic arrival matrix at time slot $t$ to CBs, and $\mathbf{D_c}(t)$ is the service rate matrix of CBs at time slot $t$. 
		Solving (\ref{tri-cb-matrix-1}) recursively as before yields:
		\begin{equation}
		\label{tri-cb-matrix-2}
		\mathbf{N_c}(t) = \mathbf{N_c}(0) +  \displaystyle\sum_{\gamma=0}^{t} \mathbf{A_c}(\gamma) - \displaystyle\sum_{\gamma=0}^{t}\mathbf{D_c}(\gamma)
		\end{equation}
		Because a CB is serviced at least once every $Nk$ time slots. The service rate of the CB at $OP(v)$ at time slot $t$, $d_{c_{v}}(t)$ is:
		\[\frac{1}{Nk} \leq d_{c_{v}}(t) \leq 1\]
		and service matrix of CBs is:
		\begin{equation}\label{eq:Dc_tri}
		\mathbf{D_c}(t)=[d_{c_{v}}(t)]
		\end{equation}
		The aggregate traffic arrival to CBs, $\mathbf{R_4}$, or:
		\begin{equation}
		\label{tri-cb-matrix-2a}
		\mathbf{R_4} = \displaystyle\sum_{\gamma=0}^{t} A_c(\gamma) 
		\end{equation}
		Let us assume the worst case scenario, where the CB is service only once in $Nk$ timeslots or  $d_{c_{v}}(t)= \frac{1}{Nk} ~ \forall ~v$ in (\ref{eq:Dc_tri}). Substituting (\ref{eq:Dc_tri}) and (\ref{tri-cb-matrix-2a}) into (\ref{tri-cb-matrix-2}) gives:
		\begin{equation}
		\label{tri-cb-matrix-3}
		\mathbf{N_c}(t) = \mathbf{N_c}(0) + \mathbf{R_4} - \frac{1}{Nk} * \vec{1}
		\end{equation}
		where
		\begin{equation}
		\label{tri-cb-matrix-4}
		\mathbf{R_4} - \frac{1}{Nk} * \vec{1} \leq \epsilon < \infty
		\end{equation}
		Because $\mathbf{R_4}$ is admissible, as discussed in Section III.A, substituting $\mathbf{R_4}$ into (\ref{tri-cb-matrix-4}) shows that $\epsilon$ is finite.
		We can conclude from  (\ref{tri-cb-matrix-3}) and (\ref{tri-cb-matrix-4}), that the occupancy of CB is also weakly stable.\\
		$\blacksquare$ \\
		This completes the proof of Theorem \ref{theorem:throughput_tri}.
		\\
		$\blacksquare$\color{black}
\noindent

\section{Analysis of In-Sequence Service}
\label{sec:proof}

In this section, we demonstrate that the TRIDENT switch forwards cells in sequence to the OPs through the proposed in-sequence forwarding mechanism.
Table \ref{tab: term2} lists the terms used in the in-sequence analysis of the proposed TRIDENT switch. 
Here, $c_{y, \tau}(i,s,j,d)$ denotes the $\tau$th cell of traffic flow $y$, which comprises cells going from $IP(i,s)$ to $OP(j,d)$. 
In addition, $t_{a_{y,\tau}}$ denotes the arrival time of $c_{y,\tau}$, and $q_{V_{y,\tau}}$ and $q_{C_{y,\tau}}$ denote the queuing delays experienced by $c_{y,\tau}$ at $VIMOQ(r,i,j,d)$ and $CB(r,j,d,i,s)$, respectively. The departure times of $c_{y,\tau}$ from the corresponding VIMOQ and CB are denoted as $d_{V_{y,\tau}}$ and $d_{C_{y,\tau}}$, respectively. 
We consider admissible traffic in this analysis.

Here, we claim that TRIDENT forwards cells in sequence to the output ports, through the following theorem.
\vspace{0.1in}
\begin{theorem}
	\label{theorem:insequence}
	{\it For any two cells $c_{y, \tau}(i,s,j,d)$ and $c_{y, \tau'}(i,s,j,d)$, where $\tau < \tau'$, $c_{y, \tau}(i,s,j,d)$ departs the destined output port before $c_{y, \tau'}(i,s,j,d)$.}
\end{theorem}

\begin{table}[htb]
	\centering
	\caption{Notations for in-sequence analysis.}
	\vskip .25in
	\label{tab: term2}
		\begin{tabular}{l | p{7cm}} \hline \hline
		$c_{y,\tau}$         & The $\tau$th cell of flow $y$ from $IP(i,s)$ to $OP(j,d)$.\\ \hline
		$t_{a_{y,\tau}}$     & Arrival time of $c_{y,\tau}$ at $IP(i,s)$.\\ \hline
		$N_{V_{y,\tau}}$     & The number of cells at $VIMOQ(r,i,j,d)$ upon the arrival of $c_{y,\tau}$.\\ \hline
		$q_{H_{y,\tau}}$     & The residual queuing delay of the HoL cell at $VIMOQ(r,i,j,d)$ upon the arrival of $c_{y,\tau}$.\\ \hline
		$q_{V_{y,\tau}}$     & Queuing delay of $c_{y,\tau}$ at $VIMOQ(r,i,j,d)$.\\ \hline
		$d_{V_{y,\tau}}$     & Departure time of $c_{y,\tau}$ from $VIMOQ(r,i,j,d)$ at $I_{C}(r,p)$.\\ \hline
		$N_{C_{y,\tau}}$      & The number of cells at  $CB(r,j,d,i,s)$ upon the arrival of $c_{y,\tau}$.\\ \hline
		$q_{C_{y,\tau}}$     & Queuing delay of $c_{y,\tau}$ at $CB(r,j,d,i,s)$ of $OP(j,d)$.\\ \hline
		$d_{C_{y,\tau}}$     & Departure time of $c_{y,\tau}$ from $CB(r,j,d,i,s)$.\\ \hline \hline
	\end{tabular}
\end{table}


\begin{lemma}  
	\label{lemma:lemma1}
	{\it For any flow traversing the TRIDENT switch, an older cell is always placed ahead of a younger cell from the same flow in the same crosspoint buffer.}
\end{lemma}

{\bf Proof:}
From the architecture and configuration of the switch an IP connects to a CM once every $k$ time slots. If a younger cell arrives at the OM before an older cell then the younger cell was forwarded through a different CM from the one the older cell was buffered. Also, two cells of the same flow may be queued in the same CB if and only if the younger cell arrived at the VIMOQ $k$ time slots later than the older cell, and therefore, the younger cell would be lined up in a queue position behind the position of the older cell. 

$\blacksquare$	\\
\\
\begin{lemma}  
	\label{lemma:lemma2}
	{\it For any number of flows traversing the TRIDENT switch, cells from the same flow are cleared from the OP in the same order they arrived at the IP.}
\end{lemma}

{\bf Proof:} 
Let us consider a traffic scenario where multiple flows are traversing the switch. We focus on one flow with cells arriving back to back. Let us also consider as an initial condition that all CBs are empty, and the VIMOQ to where the first cell of the flow is being sent has backlogged cells (from other flows) while other VIMOQs to where the subsequent cells of the same flow are sent are empty. This scenario would have the largest probability to delay the first cell of the flow and, therefore; to forward the subsequent cells of the flow out of sequence. Also, let us consider that the flow pointer at the output ports initially points to the cell arrival order $L_{y\theta}$, where $y$ is the flow id and $\theta$ is the cell's order of arrival.

Also, let us assume that the cells arrive at $L_{I}(i,r)$ one or more time slots before the configuration of the CM allows forwarding a cell to its destined OM. Thus, a cell may depart in the following or a few time slots after its arrival. This cell then may wait up to $k-1$ time slots for a favorable interconnection to take place at the CM before being forwarded to the destined OM. 
In the remainder of the discussion, we show that the arriving cells are forwarded to the destination OP in the same order they arrive in the IP.

Given flow $y$, the arrival time of the first cell $c_{y,\tau}$ is:
\begin{equation}\label{tarrForztau}
t_{a_{y,\tau}} = t_x
\end{equation}
Upon arriving in the IP, $c_{y,\tau}$ is tagged with $L_{y0}$ and forwarded to the VIMOQ. Based on the backlog condition, $c_{y,\tau}$ is placed behind $\gamma$ cells from other flows upon arriving at the VIMOQ. Therefore, the VIMOQ occupancy, $N_{V_{y,\tau}}$, is:
\begin{equation}\label{N2CzTau}
N_{V_{y,\tau}} = \gamma
\end{equation}
Using (\ref{N2CzTau}) the queuing delay of  $c_{y,\tau}$ at the VIMOQ is:
\begin{equation}\label{q2ztau}
q_{V_{y,\tau}} =  q_{H_{y,\tau}} + (\gamma - 1) k + k 
\end{equation}
where $q_{H_{y,\tau}}$ is the time it takes the HoL cell to depart the VIMOQ and $(\gamma - 1)k$ is the delay generated by the other $(\gamma - 1)$ cells ahead of $c_{y,\tau}$ in the VIMOQ.
The extra $k$ time slots are the delay $c_{y,\tau}$ experiences as it waits for the configuration pattern to repeat after the last cell ahead of it is forwarded to the OM.\\
Using (\ref{tarrForztau}) and (\ref{q2ztau}), the departure time of $c_{y,\tau}$ from the VIMOQ is:
\begin{equation}\label{d2ztau}
d_{V_{y,\tau}} = t_{a_{y,\tau}}  + q_{H_{y,\tau}} + \gamma k 
\end{equation}
When $c_{y,\tau}$ arrives at the output module it is stored at the corresponding output buffer before being forwarded to the output port.

Let us now consider the next arriving cell from flow $y$, $c_{y,\tau+ \theta}$, where $0 < \theta < k$. The time of arrival of $c_{y,\tau+ \theta}$ is:
\begin{equation}\label{taz1}
t_{a_{y,\tau + \theta}} = t_x + \theta
\end{equation}
Upon arrival, $c_{y,\tau+ \theta}$ would have $L_{y\theta}$ appended to it and forwarded to the VIMOQ.
Based on the traffic scenario, $c_{y,\tau+\theta}$ would be forwarded to an empty VIMOQ. The queuing delay at the $VIMOQ$ for $c_{y,\tau + \theta}$ is:
\begin{equation}\label{q2z1}
q_{V_{y,\tau + \theta}} = \beta
\end{equation}
where $\beta$ is the number of time slots before the configuration pattern enables forwarding $c_{y,\tau + \theta}$ to the destined OM.
Using (\ref{d2ztau}), (\ref{taz1}), and (\ref{q2z1}), the departure time of $c_{y,\tau+\theta}$ from the VIMOQ is:
\begin{equation}\label{d2z1}
d_{V_{y,\tau+\theta}} = t_x + \theta + \beta
\end{equation}

At the output port, the pointers all initially pointed to $L_{y0}$ based on the initial condition.
Therefore, irrespective of $d_{V_{y,\tau+ \theta}} < d_{V_{y,\tau}}$, for $\: \theta + \beta < q_{H_{y,\tau}} + \gamma k$, $c_{y,\tau+\theta}$ remains stored at the output buffer until $c_{y,\tau}$ is cleared from the output port, because the pointer points to $L_{y0}$.
Because CBs are empty as initial condition, the CB occupancy, $N_{C_{y,\tau}}$, upon $c_{y,\tau}$ arrival is:
\begin{equation}\label{N3}
N_{C_{y,\tau}} = 0
\end{equation}
and the occupancy of the CB, $N_{C_{y,\tau + \theta}}$, upon $c_{y,\tau + \theta}$ arrival is
\begin{equation}\label{N3_theta}
N_{C_{y,\tau + \theta}} = 0
\end{equation}
Using (\ref{N3}), the queuing delay, $q_{C_{y,\tau}}$, at the CB for $c_{y,\tau}$ is:
\begin{equation}\label{q3z}
q_{C_{y,\tau}} = 0
\end{equation}
From (\ref{d2ztau}), (\ref{d2z1}), and (\ref{N3_theta}), the queuing delay, $q_{C_{y,\tau + \theta}}$, at the CB for $c_{y,\tau+\theta}$ is:
\begin{equation}\label{q3z1}
q_{C_{y,\tau + \theta}} = q_{H_{y,\tau}} + \gamma k - \beta
\end{equation}
From (\ref{tarrForztau}), (\ref{d2ztau}), and (\ref{q3z}), the departure time of $c_{y,\tau}$ from the OP, $d_{C_{y,\tau}}$, is:
\begin{equation}\label{d3ztau}
d_{C_{y,\tau}} = t_x + 1 + q_{H_{y,\tau}} + \gamma k 
\end{equation}
From (\ref{taz1}), (\ref{d2z1}), and (\ref{q3z1}), the departure time of $c_{y,\tau+\theta}$ from the OP, $d_{C_{y,\tau+\theta}}$, is:
\begin{equation}\label{d3ztau1}
d_{C_{y,\tau+\theta}} = t_x + 1 + \theta + q_{H_{y,\tau}} + \gamma k 
\end{equation}
Using (\ref{d3ztau}) and (\ref{d3ztau1}),
\begin{equation}\label{diff}
d_{C_{y,\tau+\theta}} - d_{C_{y,\tau}} = \theta
\end{equation}
The difference between the departure times of any two cells of a flow from the CB is a function of $\theta$, which is the arrival time difference between any two cells. Therefore, cells of a flow are forwarded to the OP in the same order they arrived.

$\blacksquare$	\\

This completes the proof of Theorem \ref{theorem:insequence}.

$\blacksquare$	\\

\section{Performance Analysis}
\label{sec:simulation-performance}

We evaluated the performance of TRIDENT through computer simulation under uniform  
traffic model and compared with that of an output-queued (OQ), Space-Memory-Memory (SMM), and a Memory-Memory-Memory Clos-network (MMM) switch. We also evaluated the performance of TRIDENT through computer simulation under nonuniform traffic model and compared with that of an output-queued (OQ), space-Memory-Memory (SMM), Memory-Memory-Memory Clos-network (MMM), and MMM switch with extended memory (MM$^e$M) switches.
The SMM switch uses desynchronized static round robin at IMs and select celss from the buffers at CMs and OMs.         
The MMM switch selects cells from the buffers in the previous stage modules using forwarding arbitration schemes and is prone to serving cells out of sequence. Considering that most load-balancing switches based on Clos networks deliver low performance, we select these switches for comparison because they achieve the highest performance among Clos-network switches, despite been categorized as different architectures. We considered switches with size $N = \{64, 256\}$. For performance analysis, queues are assumed long to avoid cell losses and to identify average cell delay.

Table \ref{tab:TRIDENT_comparison} shows a comparison between the architectures of OQ, SMM, MMM, MM$^eM$, and TRIDENT.
\begin{table*}[htb]
	\centering
	\caption{Switches used in performance comparison to TRIDENT.\label{tab:TRIDENT_comparison}}
	\vskip .25in
	\begin{adjustbox}{width=\linewidth}
		\begin{tabular}{| l |p{2cm} |p{3cm} |p{3cm}| p{3cm}| p{3cm} |} \hline \hline 
			Architecture &OQ & SMM&  MMM & MM$^e$M & TRIDENT \\ \hline \hline
			Scalability&Non scalable &Scalable &Scalable &Scalable& Scalable\\ \hline
			Packet order preserved	&Yes & No & No & No & Yes\\ \hline
			Speedup & N & 1 & 1 & 1 & 1\\ \hline
			Configuration scheme & N/A & Desynchronized static round robin at IM and select cells from the buffers at CMs and OMs  & Select cells from buffers in the previous stage modules & 
			Select cells from the buffers in the previous stage modules & Prederministic and periodic \\ \hline
			On-line complexity for crossbar connections & O(1) &O(N) & O(N) & O(N) &O(1) \\ \hline
			Internal blocking & Non blocking & Blocking & Blocking & Non blocking & Non blocking \\ \hline
			Total number of VOQs per IM &N/A & $Nn$ & $Nn$ &$Nn$ & 0 \\ \hline
			Total number of Virtual central module queues per  IM & N/A & 0 &$mn$ & $nN$ & 0 \\ \hline
			Total number of virtual output (module or port) queues per CM  & N/A & $k^2$ &$k^2$ & $mN$ & $kN$ \\ \hline
			Total number of queues per OM & N/A & $mn$ & mn & $nN$ & $N^2k$ \\ \hline
			Total number of queues per OP & $N$ & 0 &0 &0 & 0 \\ \hline
		\end{tabular}
	\end{adjustbox}
\end{table*}

\subsection{Uniform Traffic}
Uniform distribution is mostly considered to be benign and the average rate for each output port $\lambda_{i,s,j,d} = \frac{1}{N}$. where $IP(i,s)$ is the source IP and $OP(j,d)$ is the destination OP. Hence, a packet arriving at the IP has an equal probability of being destined to any OP.
Figures~\ref{fig:new-tri-uniform-64} and {\ref{fig:new-tri-uniform-256} show the average under uniform traffic with Bernoulli arrivals for $N=64$ and $N=256$, respectively.
The finite and moderate average queuing delay indicated by the results shows that TRIDENT achieves 100\% throughput under this traffic pattern. This throughput is the result of the efficient load-balancing process in the IM stage. However, such high performance is expected for uniformly distributed input traffic. 

TRIDENT switch experiences a slightly higher average delay than the OQ switch. This delay is the result of cells being queued in the VIMOQs until a configuration occurs that enables forwarding the cells to their destined output modules. Due to the amount of memory required by MM$^e$M to implement the extended set of queues, our simulator can only simulate small MM$^e$M switches for queueing analysis, so we simulated the switches under this traffic pattern for $N = 64$.
This figure also shows that TRIDENT achieves a lower average delay than the MMM switch.

\begin{figure}[tb]
	\centering
	\includegraphics[width=3.0 in]{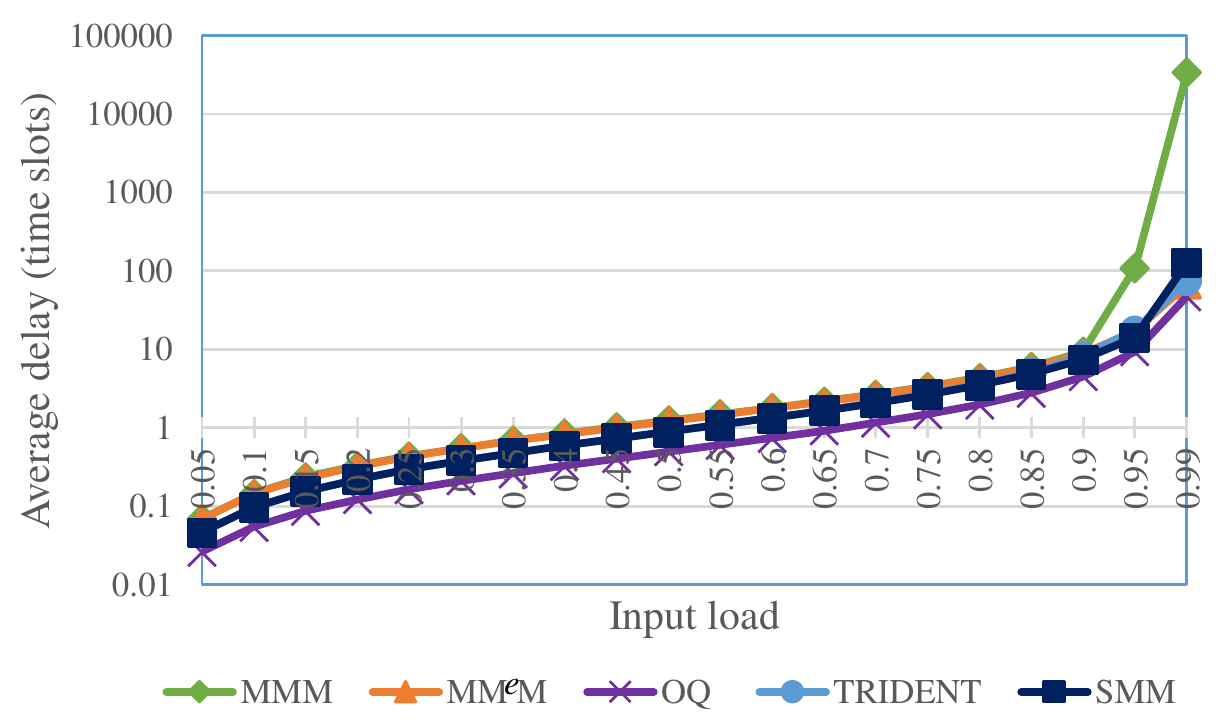}
	\caption{Average queueing delay under uniform traffic for $N$=64.}
	\label{fig:new-tri-uniform-64}
\end{figure}

\begin{figure}[tb]
	\centering
	\includegraphics[width=3.0 in]{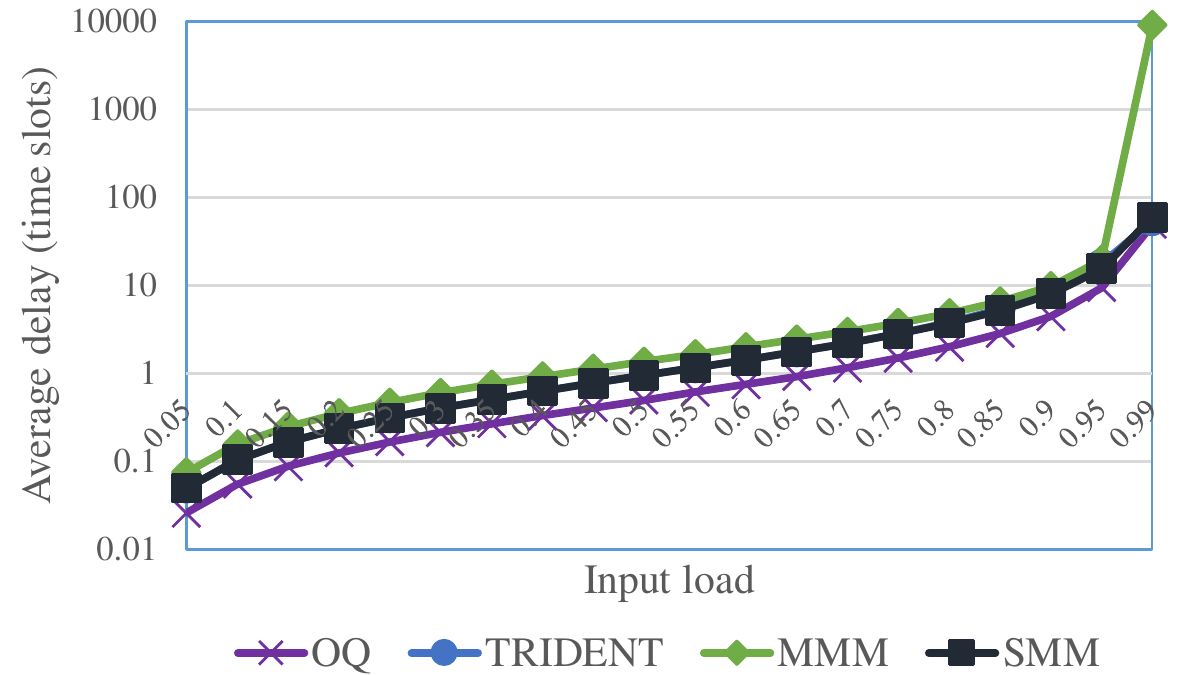}
	\caption{Average queueing delay under uniform traffic for $N$=256.}
	\label{fig:new-tri-uniform-256}
\end{figure}

Uniform bursty traffic is modeled as an ON-OFF Markov modulated process, with an average duration of the ON period set as the average burst length, $l$, with $l=\{10, 30\}$ cells.
Figures \ref{fig:new-tri-bursty10} and \ref{fig:new-tri-bursty30} show the average delay under uniform traffic with bursty arrivals for average burst length of 10 and 30 cells, respectively.
The results show that TRIDENT achieves 100\% throughput under bursty uniform traffic and it is not affected by the burst length, while the MMM switch has a throughput of 0.8 and 0.75 for an average burst length of 10 and 30 cells, respectively. Therefore, TRIDENT achieves a performance closer to that of the OQ switch.
\begin{figure}[tb]
	\centering
	\includegraphics[width=3.0 in]{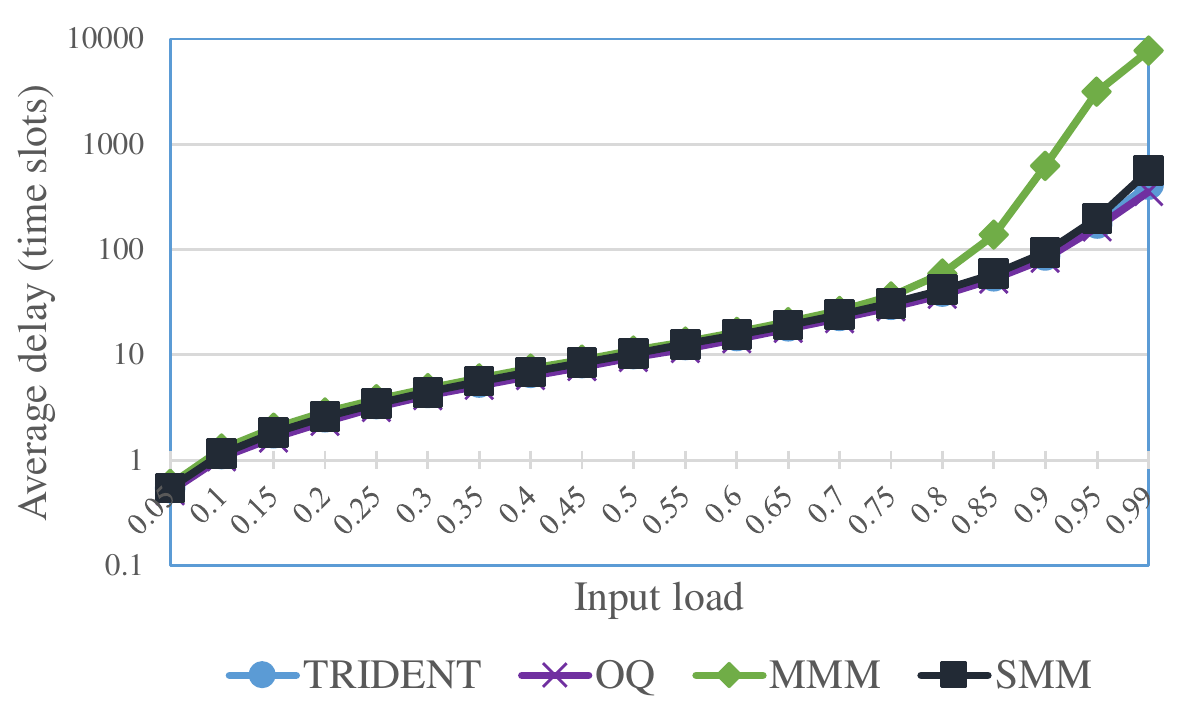}
	\caption{Average queuing delay under uniform bursty traffic with average burst length $l$=10.}
	\label{fig:new-tri-bursty10}
\end{figure}

\begin{figure}[tb]
	\centering
	\includegraphics[width=3.0 in]{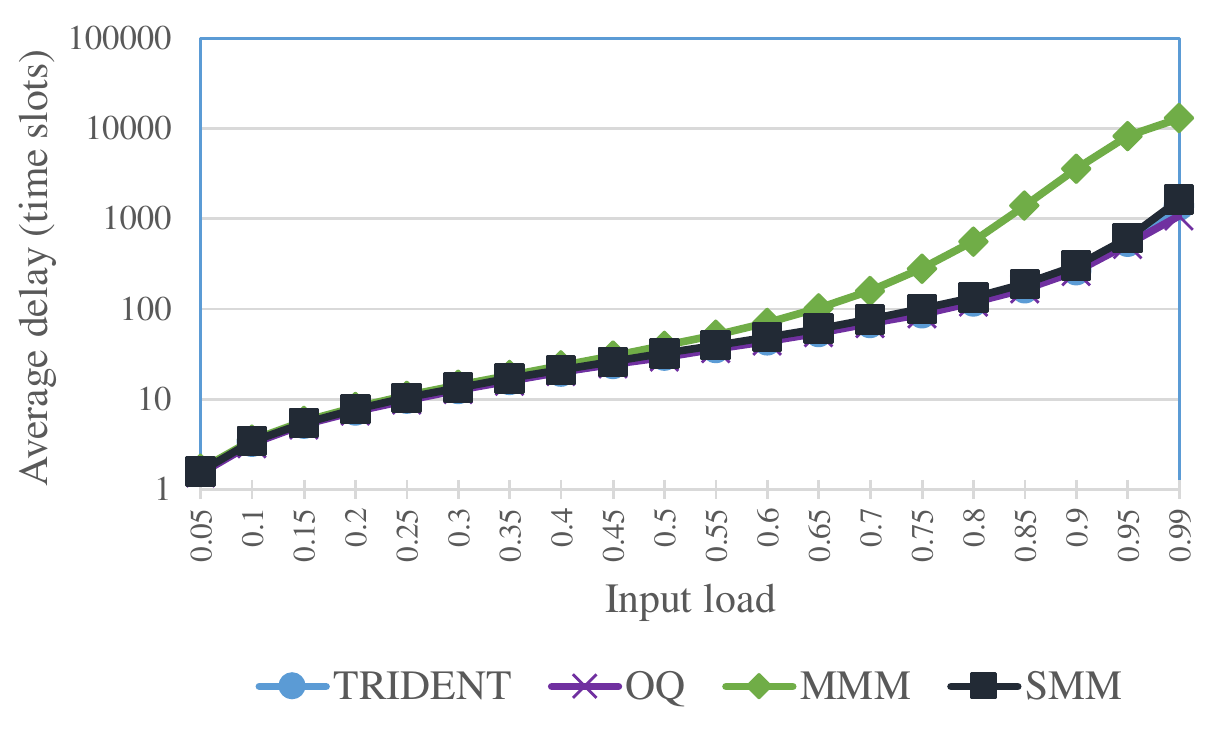}
	\caption{Average queuing delay under uniform bursty traffic with average burst length $l$=30.}
	\label{fig:new-tri-bursty30}
\end{figure}
The uniform distribution of the traffic and the load-balancing stage helps to attain this low queueing delay and high throughput. 
Figures \ref{fig:new-tri-uniform-64}, \ref{fig:new-tri-uniform-256}, \ref{fig:new-tri-bursty10}, and \ref{fig:new-tri-bursty30} show that the queueing delay difference between TRIDENT and the OQ switch is not significant. 
Figures \ref{fig:new-tri-uniform-64}, \ref{fig:new-tri-uniform-256}, \ref{fig:new-tri-bursty10}, and \ref{fig:new-tri-bursty30} also show that TRIDENT outperforms the SMM switch for all tested traffic patterns at high input load. Because the SMM switch uses load-balancing at the bufferless IMs which enables it to attain high performance similar to TRIDENT at low input load, but at high input load the configuration complexity at CMs and OMs affects its performance. In addition to the high configuration complexity required for the SMM switch as compared to TRIDENT, it also forwards cells out-of-sequence while TRIDENT forwards cells in-sequence.
 These figures also show that the effective load balancing reduces the average delay and also eliminates the offset in delay for a light load.
\subsection{Nonuniform traffic}

We also evaluated the performance of TRIDENT, MMM, MM$^e$M, and OQ switches under nonuniform traffic. We adopted the unbalanced traffic model \cite{cixb,cixob} as a nonuniform traffic pattern.  The nonuniform traffic can be modeled using an unbalanced probability $\omega$ to indicate the load variances for different flows. Consider input port $IP(i,s)$ and output port $OP(j,d)$ of the TRIDENT switch, the traffic load is determined by

\begin{equation}\label{equ25}
\rho_{i,s,j,d} = 
\begin{dcases}
\rho(\omega + {\frac{1 - \omega}{N}}),& \text{if} ~i = j ~\text{and}~s = d,\\
\rho{\frac{1 - \omega}{N}},              & \text{otherwise}
\end{dcases}
\end{equation}    
where $\rho$ is the input load for input $IP(i,s)$ and $\omega$ is the unbalanced probability.
When $\omega$=0, the input traffic is uniformly distributed and when $\omega$=1, the input traffic is completely directional; traffic from $IP(i,s)$ is destined for $OP(j,d)$.

Figure \ref{fig:new-tri-unb-throughput} shows the throughput of TRIDENT, SMM, MMM, and MM$^e$M switches. The figure shows that TRIDENT switch attains 100\% throughput under this traffic pattern for all values of $\omega$, matching the performance of SMM and MM$^e$M and outperforming that of MMM. These three buffered switches are known to achieve high throughput at the expense of out-of-sequence forwarding.

We also tested the average queueing delay of TRIDENT under this nonuniform traffic. It has been shown that many switches do not achieve high throughput when $\omega$ is around 0.6 \cite{cixob}. Therefore, we measured the average delay of TRIDENT under this unbalanced probability, as Figures \ref{fig:new-tri-unb-delay-64} and \ref{fig:new-tri-unb-delay-256} show for $N=64$ and $N=256$, respectively, and compared it with MMM, SMM, MM$^e$M, and OQ switches.  One should note that due to the limited scalability of MMM and MM$^e$M, the comparison of TRIDENT for $N=256$ under this traffic conditions only includes SMM and OQ switches. Figure \ref{fig:new-tri-unb-delay-256} shows that the delay of TRIDENT is lower than the delay achieved by SMM under high input loads.

As Figure \ref{fig:new-tri-unb-delay-64} for $N=64$ shows, the average delay of TRIDENT is lower than the delay achieved by SMM, MMM, and MM$^e$M under high input loads while also achieving a comparable delay of an OQ switch. The small performance difference between TRIDENT and OQ is similar for $N=256$, as Figure \ref{fig:new-tri-unb-delay-256} shows. These results are achieved because the load-balancing stage of TRIDENT distributes the traffic uniformly throughout the switch. Therefore, the queuing delay is similar to that observed under uniform traffic. These results also show that high switching performance of TRIDENT is not affected by the in-sequence mechanism of the switch and the load-balancing effect is more noticeable under nonuniform traffic.

\begin{figure}[tb]
	\centering
	\includegraphics[width=3.0 in]{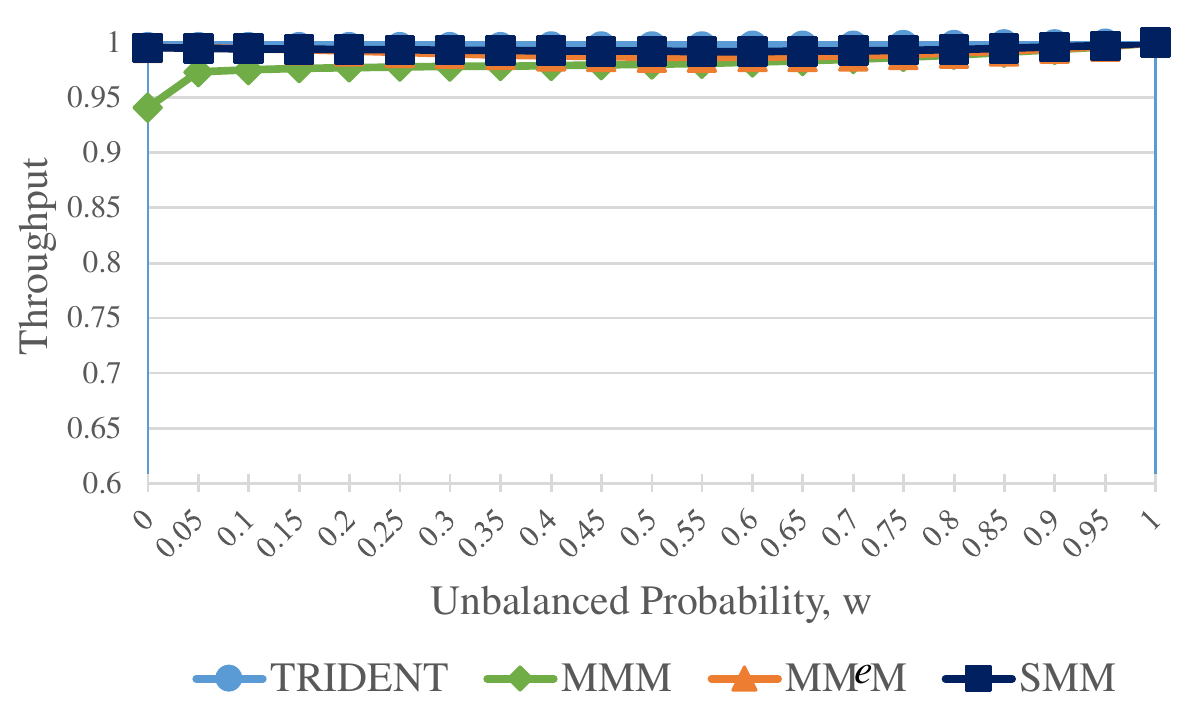}
	\caption{Throughput under unbalanced traffic for $0 \leq w \leq 1.0$ and $N$=256.}
	\label{fig:new-tri-unb-throughput}
\end{figure}

\begin{figure}[tb]
	\centering
	\includegraphics[width=3.0 in]{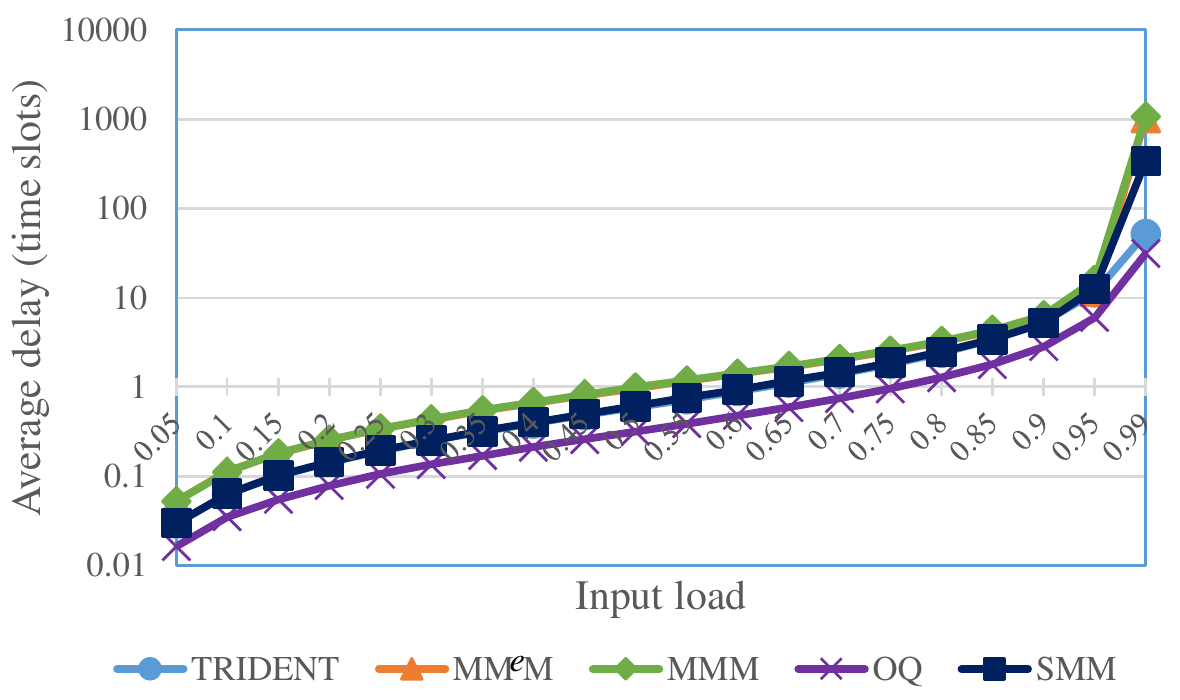}
	\caption{Average queuing delay under unbalanced traffic with $w=0.6$ for $N$=64.}
	\label{fig:new-tri-unb-delay-64}
\end{figure}

\begin{figure}[tb]
	\centering
	\includegraphics[width=3.0 in]{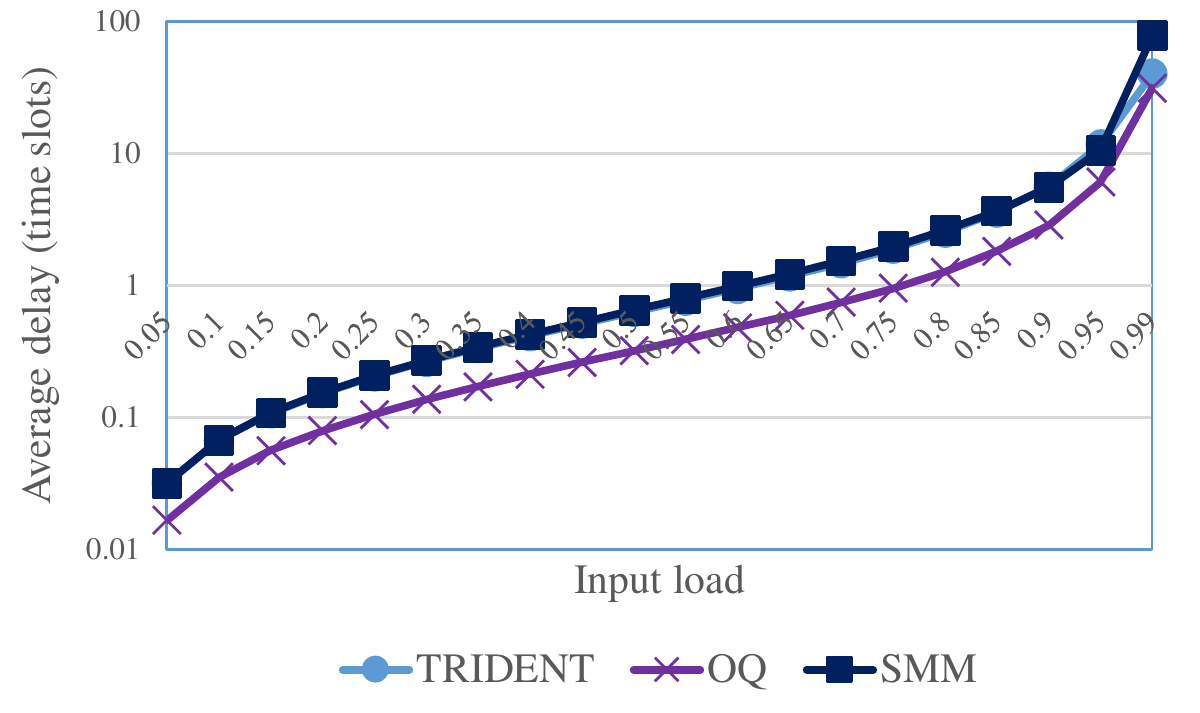}
	\caption{Average queuing delay under unbalanced traffic with $w=0.6$ for $N$=256.}
	\label{fig:new-tri-unb-delay-256}
\end{figure}
In addition to the analysis in \ref{proof-cb}, we also tested the impact of the CB size through computer simulations. Where we tested and measured the average delay under unbalanced traffic and throughput under hot-spot per port traffic models, for three TRIDENT switches with CB sizes of $k^2$, $N^2$, and $\infty$, respectively. Figure \ref{fig:TRIDENT_queue_compare} shows that the size of the crosspoint buffer does not impact the switch performance.
The TRIDENT switches, each with different crosspoint buffer size, attains 100\% throughput for hotpsot per port traffic model. Which also indicates that the size of the CB does not impact the performance of the switch as shown in the analysis above.
where TRIDENT short-queue has a crosspoint buffer size of $k^2$, TRIDENT short-queue has a crosspoint size of $N^2$, and TRIDENT infinite-queue has an infinite crosspoint buffer size.
\begin{figure}[tb]
	\centering
	\includegraphics[width=3.0 in]{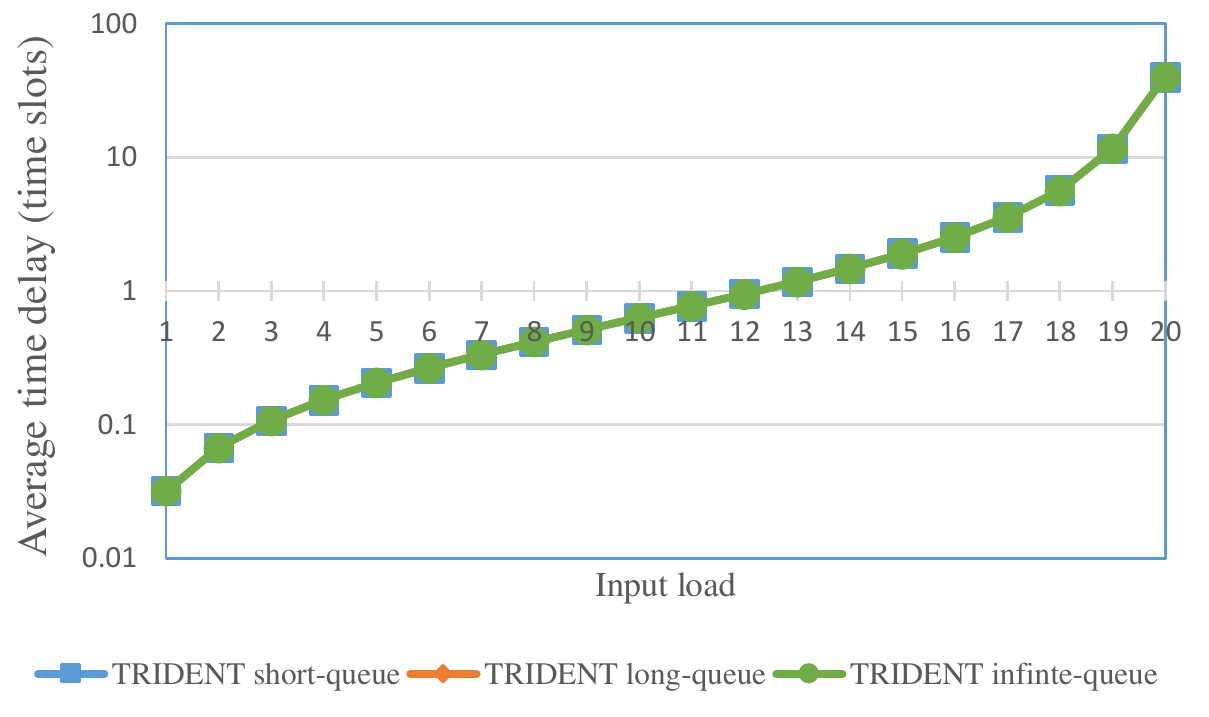}
	\caption{Average queuing delay under unbalanced traffic with $w=0.6$ for $N$=256.}
	\label{fig:TRIDENT_queue_compare}
\end{figure}

\section{Conclusions}
\label{sec:conclusions}

We have introduced a three-stage load-balancing packet switch that has virtual output module queues between the input and central stages, and a low-complexity scheme for configuration and forwarding cells in sequence for this switch. We call this switch TRIDENT. To effectively perform load balancing TRIDENT has virtual output module queues between the IM and CM stages. Here, IMs and CMs are bufferless modules, while the OMs are buffered ones.
All the bufferless modules of TRIDENT follow a predetermined configuration while the OM selects the cell of a flow to be forwarded to an output port based on the cell's arrival order and uses round-robin scheduling to select the flow to be served. Because of the buffers at crosspoints of OMs, the switch rescinds port matching, and the configuration complexity of the switch is minimum, making it comparable to that of MMM switches. We introduce an in-sequence mechanism that operates at the outputs based on arrival order inserted at the inputs of TRIDENT to avoid out-of-sequence forwarding caused by the central buffers. We modeled and analyzed the operations of each of the stages and how they affect the incoming traffic to obtain the loads seen by the output ports. We show that for admissible independent and identically distributed traffic, the switch achieves 100\% throughput. This high performance is achieved without resorting to speedup nor switch expansion. In addition, we analyzed the operation of the forwarding mechanism and demonstrated that it forwards cells in sequence. We showed, through computer simulation, that for all tested traffic, the switch achieves 100\% throughput for uniform and nonuniform traffic distributions.

\bibliographystyle{IEEEtran}
\bibliography{Reference}

\end{document}